\documentclass[a4paper,11pt]{article}
\usepackage{graphicx,amssymb,amstext,amsmath}
\usepackage{color, yfonts, tcolorbox}
\usepackage{floatflt}

\usepackage{float}
\usepackage{esint}
\usepackage{subcaption}
\usepackage{cite}
\usepackage{mathtools}
\usepackage{commath}
\usepackage{array}
\usepackage{boldline}
\usepackage{multirow}
\usepackage{arydshln}
\usepackage{bigdelim}
\usepackage{hyperref}
\usepackage{simpler-wick}

\definecolor{cbl}{rgb}{0,0,1}                

\newcommand{\bc}{\begin{center}}
\newcommand{\ec}{\end{center}}
\def\ba#1{\begin{array}{#1}\displaystyle}
\newcommand{\ea}{\end{array}}

\newcommand{\beq}{\begin{equation}}
\newcommand{\eeq}{\end{equation}}
\newcommand{\beqa}{\begin{eqnarray}}
\newcommand{\eeqa}{\end{eqnarray}}

\newcommand{\bi}{\begin{itemize}}
\newcommand{\ei}{\end{itemize}}

\newcommand{\bra}{\langle}
\newcommand{\ket}{\rangle}

\newcommand{\Tr}{{\rm Tr}}

\def \be {\begin{equation}} 
\def \ee {\end{equation}} 
\def \l {\left(} 
\def \r {\right)} 
\def \la {\langle} 
\def \ra {\rangle}

\begin{document}
\begin{titlepage}
\vspace{0.2cm}
\begin{center}

{\large{\bf{Symmetry Resolved Entanglement of Excited States in Quantum Field Theory II: Numerics, Interacting Theories and Higher Dimensions}}}

\vspace{0.8cm} 
{\large Luca Capizzi{\LARGE $^{\star}$},  Cecilia De Fazio$^\clubsuit$, Michele Mazzoni$^\spadesuit$,\\ Luc\'ia Santamar\'ia-Sanz$^\diamondsuit$, and Olalla A. Castro-Alvaredo$^\heartsuit$}

\vspace{0.8cm}
{\small
{\LARGE $^{\star}$}  SISSA and INFN Sezione di Trieste, via Bonomea 265, 34136 Trieste, Italy\\
\medskip

$^{\clubsuit}$ School of Physics and Astronomy, University of Nottingham, Nottingham, NG7 2RD, UK\\
\medskip
$^{\spadesuit,\heartsuit}$ Department of Mathematics, City, University of London, 10 Northampton Square EC1V 0HB, UK\\
\medskip

\hspace{-0.5cm}$^\diamondsuit$ Departamento de F\'isica Te\'orica, At\'omica y \'Optica, Universidad de Valladolid, 47011 Valladolid, Spain\\

}
\end{center}

\medskip
\medskip
\medskip
\medskip

 In a recent paper we studied the entanglement content of zero-density excited states in complex free quantum field theories, focusing on the symmetry resolved entanglement entropy (SREE). By zero-density states we mean states consisting of a fixed, finite number of excitations above the ground state in an infinite-volume system. The SREE is defined for theories that possess an internal symmetry and provides a measure of the contribution to the total entanglement of each symmetry sector. In our work, we showed that the ratio of Fourier-transforms of the SREEs (i.e. the ratio of charged moments) takes a very simple and universal form for these states, which depends only on the number, statistics and symmetry charge of the excitations as well as the relative size of the entanglement region with respect to the whole system's size. In this paper we provide numerical evidence for our formulae by computing functions of the charged moments in two free lattice theories: a 1D Fermi gas and a complex harmonic chain. We also extend our results in two directions: by showing that they apply also to excited states of interacting theories (i.e. magnon states) and by developing a higher dimensional generalisation of the branch point twist field picture, leading to results in (interacting) higher-dimensional models. 

\noindent 
\medskip
\medskip
\medskip
\medskip

\noindent {\bfseries Keywords:} Quantum Entanglement, Symmetry Resolved Entanglement, Integrable Quantum Field Theory, Excited States,  Branch Point Twist Fields

\vfill
\noindent 
{\LARGE $^{\star}$} lcapizzi@sissa.it\\
$^\clubsuit$ cecilia.defazio@nottingham.ac.uk\\
$^\spadesuit$ michele.mazzoni.2@city.ac.uk\\
$^\diamondsuit$ lucia.santamaria@uva.es\\
{$^\heartsuit$}o.castro-alvaredo@city.ac.uk

\hfill \today

\end{titlepage}
\section{Introduction}
The symmetry resolved entanglement entropy (SREE) is a relatively new measure of entanglement for many-body quantum systems. 
The concept was introduced in \cite{Caputa, newHo,newHoo} although the name SREE first appeared in \cite{GS}, where it was defined as a new measure of entanglement associated to theories that posses an internal discrete or continuous symmetry. The role of symmetries in the structure of the entanglement entropy and the contributions of symmetry sectors to the total entropy were also studied in \cite{german3}, simultaneously and independently of \cite{GS}. The approach of \cite{GS}, which we adopt here, falls within the framework of branch point twist fields \cite{kniz,orbifold,Calabrese:2004eu,entropy,benext} and their composite generalisations \cite{ctheorem,Levi,BCDLR,GS}.
These fields have only been formally defined for 1+1D theories and have proven particularly useful in the study of entanglement measures for conformal field theories (CFTs) \cite{Calabrese:2004eu,negativity1,disco1,paola} and integrable quantum field theories \cite{entropy,back,Isingb,E8,FB}. In this context, typical examples of theories that posses internal symmetries are the Ising field theory, with (discrete) $\mathbb{Z}_2$-symmetry, and the sine-Gordon model which has (continuous) $U(1)$-symmetry. 

Interest in the SREEs has been motivated in part by the fact that they can be measured experimentally  \cite{expSRE1,expSRE2,expSRE3}. Our interest however stems from 1+1D quantum field theory and related models, such as integrable spin chains, where powerful analytical and numerical computational techniques can be employed. In this context, there has been an enormous wealth of results in recent years. These include many studies in CFT \cite{GS,Laflorencie_2014,german3,FG,Bonsignori_2020,Capizzi_2020,Murciano_2021,Chen_2021,capizzi2021,dubailmur,EID, Chen1}, integrable quantum field theories \cite{Bonsignori_2019,Murciano_2020,Horvath_2021, FFSRE,horvath2021branch,pottsSRE}, holography \cite{Caputa, newHo,newHoo, Zhao_2021,weisenberger2021,meyer}, lattice models \cite{Laflorencie_2014,german3,FG,Bonsignori_2019,Bonsignori_2020,Fraenkel_2020,BHM,BCM,MDGC,CCDGM,PBC,TR,MRC2}, out of equilibrium  \cite{FG, PBC, expSRE2,fraenkel2021,parez2021exact,parezlatest, Chen2} and for systems with unusual dynamics \cite{TRVC,KUFS,KUFS2,operate,CLSS,tophases}. 

Let us recall the general definition of the SREEs. Consider a 1+1D quantum field theory and a bipartition of space into two complementary regions $A$ and $\bar{A}$ so that the Hilbert space of the theory $\mathcal{H}$ also decomposes into a direct product $\mathcal{H}_A \otimes \mathcal{H}_{\bar{A}}$. Assume that the theory is in a pure state $|\Psi\ket$. We define the reduced density matrix of subsystem $A$ as
\beq
\rho_A=\mathrm{Tr}_{\bar{A}}(|\Psi\ket \bra \Psi|)\,,
\eeq
and the standard von Neumann and $n$th R\'enyi entropy of subsystem $A$ are defined as
\beq
S=-\mathrm{Tr}_A(\rho_A \log \rho_A)\quad \mathrm{and} \quad S_n=\frac{\log(\mathrm{Tr}_A \rho_A^n)}{1-n}\quad \mathrm{with} \quad S=\lim_{n\rightarrow 1} S_n\,,
\label{SS}
\eeq
where $\mathrm{Tr}_A \rho_A^n:={\mathcal{Z}}_n/{\mathcal{Z}}_1^n$ is the normalized partition function of a theory constructed from $n$ non-interacting copies or replicas of the original model.  

Assume now that the theory has an internal symmetry, with symmetry operator $Q$ whose projection onto subsystem $A$ we call $Q_A$. By construction, we have that $[Q_A, \rho_A]=0$. Let  $q$ be the eigenvalue of operator $Q_A$ in a particular symmetry sector. Then
\beq
{\mathcal{Z}}_n(q)=\mathrm{Tr}_A(\rho_A^n \mathbb{P}(q))\,, \label{par1}
\eeq 
with $\mathbb{P}(q)$ the projector on the symmetry sector of charge $q$, is the symmetry resolved partition function. In terms of this object, the SREEs can be written as
\beq
S_n(q)=\frac{1}{1-n} \log\frac{{\mathcal{Z}}_n(q)}{{\mathcal{Z}}_1^n(q)}\quad \mathrm{and} \quad S(q)=\lim_{n\rightarrow 1} S_n(q)\,.
\label{sym}
\eeq
In particular, for $n=1$ the partition function (\ref{par1}) represents the probability of obtaining the value $q$ upon measuring the charge.

In part I of this series \cite{partI}, we presented a study of these SREEs by focusing on complex free theories and qubit states. We considered a class of excited states $|\Psi\ket$ consisting of a finite number of excitations and a particular scaling limit, namely one in which parts $A$ and $\bar{A}$ are infinitely large, but the ratio of their lengths by the total length of the system is $r$ and $1-r$ respectively, with $r\in [0,1]$.  Since the ratio of particle number to volume is zero, we call these states zero-density excited states.

As discussed in \cite{GS} the partition function (\ref{par1}) can best be obtained in terms of its Fourier modes, the  charged moments
\beq 
Z_n(\alpha)=\mathrm{Tr}_A(\rho_A^n e^{2\pi i\alpha Q_A})\,.
\label{CM}
\eeq 
If the symmetry is $U(1)$ the following relationship holds
\beq
\mathcal{Z}_n(q)=\int_{-\frac{1}{2}}^{\frac{1}{2}} d\alpha \, Z_n(\alpha) e^{-2\pi i\alpha q}\,,
\label{moments}
\eeq
which we employed in \cite{partI} in order to compute the SREEs. However, similar formulae hold for other symmetries and, in particular, the integral is replaced by a sum if the symmetry is discrete (see e.g. \cite{pottsSRE}).

Let us now summarise the main results of \cite{partI}. We computed the ratio of charge moments between the excited and ground states. As anticipated, the results are functions of the state $|\Psi\ket$, the parameter $\alpha$ and the ratio $r$. Formally, we define the ratios
\beq
M_n^\Psi(r;\alpha):=\frac{Z_n^\Psi(r; \alpha)}{Z_n^0(r; \alpha)}\,,
\label{eq_Mratio}
\eeq
 generalising the notation in the definition (\ref{CM}) to include the dependence in $r$.

\medskip

We found that for a state consisting of a single particle excitation of charge $\epsilon = \pm 1$ above the ground state $|\Psi\ket=|1^\epsilon \ket$, the ratio is
\beq
M_n^{1^\epsilon}(r;\alpha)=e^{2\pi i \epsilon \alpha} r^n + (1-r)^n\,,
\label{una}
\eeq
and for a state of two excitations with opposite charges $\pm \epsilon$ we have
\beq
M_n^{1^\epsilon 1^{-\epsilon}}(r;\alpha)=(e^{2\pi i \epsilon \alpha} r^n + (1-r)^n)(e^{-2\pi i \epsilon \alpha} r^n + (1-r)^n) \,.
\label{dos}
\eeq
For a state of $k$ identical excitations, that is, with the same charges $\epsilon$ and momenta, we obtain
\beq
M_n^{k^\epsilon}(r;\alpha)= \sum_{j=0}^k [f_j^k(r)]^n e^{ 2\pi i \epsilon j \alpha}\,,
\label{for1}
\eeq
where $ f_j^k(r):={}_kC_{j} \, r^j (1-r)^{k-j}$ and ${}_kC_{j}=\frac{k!}{j!(k-j)!}$ is the binomial coefficient. This formula acts as the fundamental building block for all other cases. A generic state comprising $s$ groups of $k_i^{\epsilon_i}$ identical particles of charge $\epsilon_i$ will have 
\beq
M_n^{k_1^{\epsilon_1}\ldots k_s^{\epsilon_s}}(r;\alpha)=\prod_{i=1}^s M_n^{k_i^{\epsilon_i}}(r;\alpha)\,.
\label{general}
\eeq
In the present follow up paper we test these results numerically as well as extend them to interacting and higher dimensional theories. 
\medskip

Note that the possibility of having identical excitations is excluded for fermionic theories.
For $\alpha=0$ these formulae reduce to those found in \cite{excited,excited2}, later generalised to entanglement measures of multiple disconnected regions \cite{excited3} and to higher dimensions for free bosons in \cite{excited4}. These results in turn have been extended in a series of works \cite{ali1,ali2,ali3,ali4,ali5,ali6} to deal with finite volume corrections and non-localised excitations. More recently, some of the $\alpha=0$ results were recovered as a semiclassical limit in the presence of an interaction potential \cite{mussvit}. This semiclassical picture had already been invoked much earlier, see for instance \cite{excitedalba}. However, it is worth emphasising that our formulae are not merely semiclassical limits but hold for genuine quantum theories. The quantum nature of the model is encapsulated by the symmetry resolved entanglement entropy of the ground state (and its associated moments), which are highly non-trivial for quantum models. In other words, it is only the ratios (\ref{eq_Mratio}) that are simple, not the individual charged moments.

\medskip
This paper is organized as follows: In Section \ref{numeros} we present numerical results for two (free) discrete systems: a 1D lattice Fermi gas and a 1D (complex) harmonic chain. We find that in both cases the formulae above are reproduced with great precision, even if the scaling limit of the Fermi theory is a massless free fermion whereas for the harmonic chain it is a complex massive boson. For the complex free boson, we show how the SREE can be computed by employing a wave-functional method partly presented in this section with technical details in Appendix A. In Section \ref{magnon} we show that our formulae also hold for some magnon states in interacting theories, including those of two interacting magnons.
In Section \ref{higher} we present an extension of our results to higher-dimensional quantum field theories, which draws on the connection between entanglement measures and twist fields and is close in spirit to the wave-functional approach employed for the complex free boson. We conclude in Section \ref{conclusion}.

\section{Numerical Results}
\label{numeros}
In this section we present numerical results for two very different discrete models. First we consider a 1D lattice Fermi gas, which has critical features but also possesses highly excited states whose entanglement is well described by our formulae, and then we look at the harmonic chain, whose scaling limit is a massive free boson. Whereas for the first model we can only consider distinct excitations, for the second we consider also states of identical excitations. The good agreement found confirms the more general picture put forward in \cite{excited} that these kinds of formulae hold under the broad assumption of localised excitations. These are present both in gapped systems due to finite mass scale/correlation length, and in critical systems, when the De Broglie wave length of the excitations (which is inversely proportional to their momentum) is sufficiently small compared to subsystem size.

\subsection{1D Lattice Fermi Gas}

In this section we analyse a particle-hole excited state of a 1D lattice Fermi gas, comparing our analytical predictions with the numerical data. Even though the model is critical, it was realised in \cite{ali3} that certain highly energetic quasiparticle excitations still have a universal entanglement content. More precisely, if one assumes that a set of quasiparticles with small enough De Broglie wavelengths (compared to the typical geometric lengths) is present and their momenta are sufficiently separated, then the quasiparticles will be essentially uncorrelated with each other and with respect to zero-point fluctuations. We refer the interested reader to \cite{ali1,ali2,ali3,ali4,ali5,ali6} for further details about the universal entanglement content of quasiparticles in critical systems.

Our goal here is to briefly review the numerical techniques involved in the characterisation of fermionic Gaussian states  \cite{Peschel-03} and their application to the computation of symmetry resolved measures. We start by considering the Hamiltonian of free spinless fermions on a circle of length $L$
\be
    H = -\frac{1}{2}\sum_j f^\dagger_{j+1}f_j+f^\dagger_{j}f_{j+1}+\mu \sum_j f_j^\dagger f_j,
\label{FFham}    
\ee
where  $\mu$ is the chemical potential and $\{f_j\}_{j=1,\dots,L}$ , $\{f_j^\dagger\}_{j=1,\dots,L}$ are the ladder operators obeying the standard anticommutation relations
\be
\{f_j,f_{j'}\} = \{f_j^\dagger ,f^\dagger _{j'}\} = 0, \qquad \{f_j,f^\dagger_{j'}\} = \delta_{jj'}.
\ee
When $|\mu|<1$ the theory is gapless, and the ground state is a Fermi sea with Fermi momentum $k_F = \text{arccos}(\mu)$. The two-point function evaluated in the ground-state at Fermi momentum $k_F$ takes the following form
\be
C_0(j,j')\equiv \la f_j^\dagger f_{j'}\ra_{0} = \frac{\sin k_F(j-j')}{L \sin \frac{\pi(j-j')}{L}}.
\label{Corr_gs}
\ee
Here, we analyse the quasiparticle excited state described by the following two-point function
\be
C(j,j') = C_0(j,j') +\frac{1}{L}e^{-i(k_F +\frac{\pi}{4}-\frac{\pi}{L})(j-j')} - \frac{1}{L}e^{ -i(k_F -\frac{\pi}{4}+\frac{\pi}{L})(j-j')}\,.
\label{Corr_exc}
\ee
It corresponds to the insertion of a fermion of momentum $k = k_F+\frac{\pi}{4}$ above the ground state and the removal of another fermion (or equivalently, the insertion of a hole) at $k = k_F -\frac{\pi}{4}$. The choice of the momentum shift $|k-k_F| = \frac{\pi}{4}$ is not important in the continuum limit, where the only necessary condition is that $|k-k_F|$ remains finite when $L\rightarrow \infty$ \footnote{In the work \cite{Capizzi_2020} another particle-hole state satisfying $|k-k_F| \sim 1/L$ was analysed. The entanglement measures of that low-lying state turned out to be captured instead by CFT predictions, due to the strong correlation effects between the particle/hole and the zero-point fluctuations.}.
We now have to specify the symmetry of the model. The Hamiltonian \eqref{FFham} is invariant under an internal $U(1)$ symmetry associated to the number of fermions generated by
\be
Q = \sum_j f^\dagger_jf_j,
\ee
and it clearly satisfies the locality condition $Q = Q_A + Q_{\bar{A}}$, with
\be
Q_A = \sum_{j \in A} f^\dagger_jf_j, \quad Q_{\bar{A}} = \sum_{j \in {\bar{A}}} f^\dagger_jf_j.
\ee
As subsystem $A$ we consider the segment of length $\ell$, that is the sites $j=1,\ldots, \ell$
and investigate its entanglement properties with the complementary region ${\bar{A}}$ containing sites $j=\ell+1,\ldots, L$. We denote by $C^A_0$ and $C^A$  the $\ell\times \ell$ matrices resulting from projection of the matrices $C_0$ and $C$ (defined by Eqs.~\eqref{Corr_gs} and \eqref{Corr_exc} respectively) onto subsystem $A$, keeping only $j=1,\dots ,\ell$ as spacial indices. Following \cite{Capizzi_2020} we express the charged moments of the particle-hole state and the ground state by means of the determinants
\be
\text{Tr}_A(\rho_A^n e^{2\pi i \alpha Q_A}) = \det\l (C^A)^n e^{2\pi i\alpha} +(1-C^A)^n\r,
\label{det_exc}
\ee
\be
\text{Tr}_A(\rho_{A,0}^n e^{2\pi i \alpha Q_A}) = \det\l (C^A_0)^ne^{2\pi i\alpha} +(1-C^A_0)^n\r,
\label{det_gs}
\ee
with $\rho_A$ and $\rho_{A,0}$ the respective reduced density matrices. According to our analytical predictions, we expect that the ratio of the charged moments takes the following universal form
\be
\frac{\text{Tr}_A(\rho_A^n e^{2\pi i \alpha Q_A})}{\text{Tr}_A(\rho_{A,0}^n e^{2\pi i \alpha Q_A})} \simeq  (r^n e^{2\pi i\alpha}+(1-r)^n)(r^n e^{-2\pi i\alpha}+(1-r)^n)\,,
\label{ch_mom_ratio}
\ee
that is, the expression for two distinct excitations with charges $\pm 1$, which contribute to the charged moment with an Aharonov-Bohm phase $e^{\pm 2\pi i\alpha}$. We write $\simeq$ to indicate that equality is only expected in the scaling limit of the lattice model.

To test the validity of Eq. (\ref{ch_mom_ratio}) we consider two entanglement measures, namely the excess of (total) R\'enyi entropy and the so-called (following the terminology of \cite{Capizzi_2020}) ``excess of variance". The excess of entropy is recovered from our formulae for $\alpha=0$ and for two distinct excitations takes the simple form 
\be
\Delta S_n \equiv \frac{1}{1-n}\log \frac{\text{Tr}_A(\rho_{A}^n)}{\text{Tr}_A(\rho_{A,0}^n)} \simeq \frac{\log \l r^n + (1-r)^n\r^2}{1-n}\,.
\label{delta_Sn}
\ee
We define the variance\footnote{The choice of this terminology comes from the fact that for $n=1$ the physical variance of the charge is obtained. For $n>1$ this variance has not a direct physical meaning, nevertheless it is still useful for the understanding of the symmetry resolved entanglement.} associated to $\rho_A$ as
\be
\la \Delta Q_A^2\ra_{n} \equiv \frac{\text{Tr}_A(\rho_{A}^n Q_A^2)}{\text{Tr}_A(\rho_{A}^n )}-\l\frac{\text{Tr}_A(\rho_{A}^n Q_A)}{\text{Tr}_A(\rho_{A}^n )}\r^2 = \frac{1}{(2\pi i)^2}\frac{d^2}{d\alpha^2}\left.\log \frac{\text{Tr}_A(\rho_{A}^n e^{2\pi i\alpha Q_A})}{\text{Tr}_A(\rho_{A}^n )}\right|_{\alpha = 0}.
\ee
Similarly, we denote by $\la \Delta Q_A^2\ra_{n,0}$ the variance of the ground state $\rho_{A,0}$. From \eqref{ch_mom_ratio} it then follows that the excess of variance is given by
\be
\la \Delta Q_A^2\ra_{n} - \la \Delta Q_A^2\ra_{n,0} \simeq \frac{2r^n(1-r)^n}{(r^n+(1-r)^n)^2}.
\label{Var_exc}
\ee
A way to physically interpret the result of Eq. \eqref{Var_exc} is to regard this excess of variance as twice the contribution associated to a single quasiparticle, since particles and antiparticles contribute in the same way. The latter is just the variance of a Bernoulli random variable with success probability given by
\be
p =\frac{r^n}{r^n+(1-r)^n},
\ee
namely the probability one associates to the presence of a quasiparticle in $A$ computed with the density matrix $\rho_A^n$. Since the variance of a Bernoulli variable with probability $p$ is just $p(1-p)$, we get Eq. \eqref{Var_exc}.

In Fig.~\ref{fig_exc} we report the numerical values of $\Delta S_n$ and $\la \Delta Q_A^2\ra_{n} - \la \Delta Q_A^2\ra_{n,0}$, computed from $\eqref{det_exc}$ and $\eqref{det_gs}$ using exact diagonalisation of the correlation matrices $C_A,C_{A,0}$, and our analytical predictions. We keep $L$ fixed, analysing different values of $r= \ell /L$. Our choice is motivated by the expectation that these plots should be \lq\lq universal\rq\rq at large $L$, meaning different data obtained with different $L$ should collapse to the same universal prediction (independent of lattice details as $k_F$) when $L\rightarrow \infty$. As we see from the plots in Fig.~\label{fig_exc}, the match between numerics and analytics is really good.

\begin{figure}[t]
	\begin{subfigure}{.5\textwidth}
	\includegraphics[width=\linewidth]{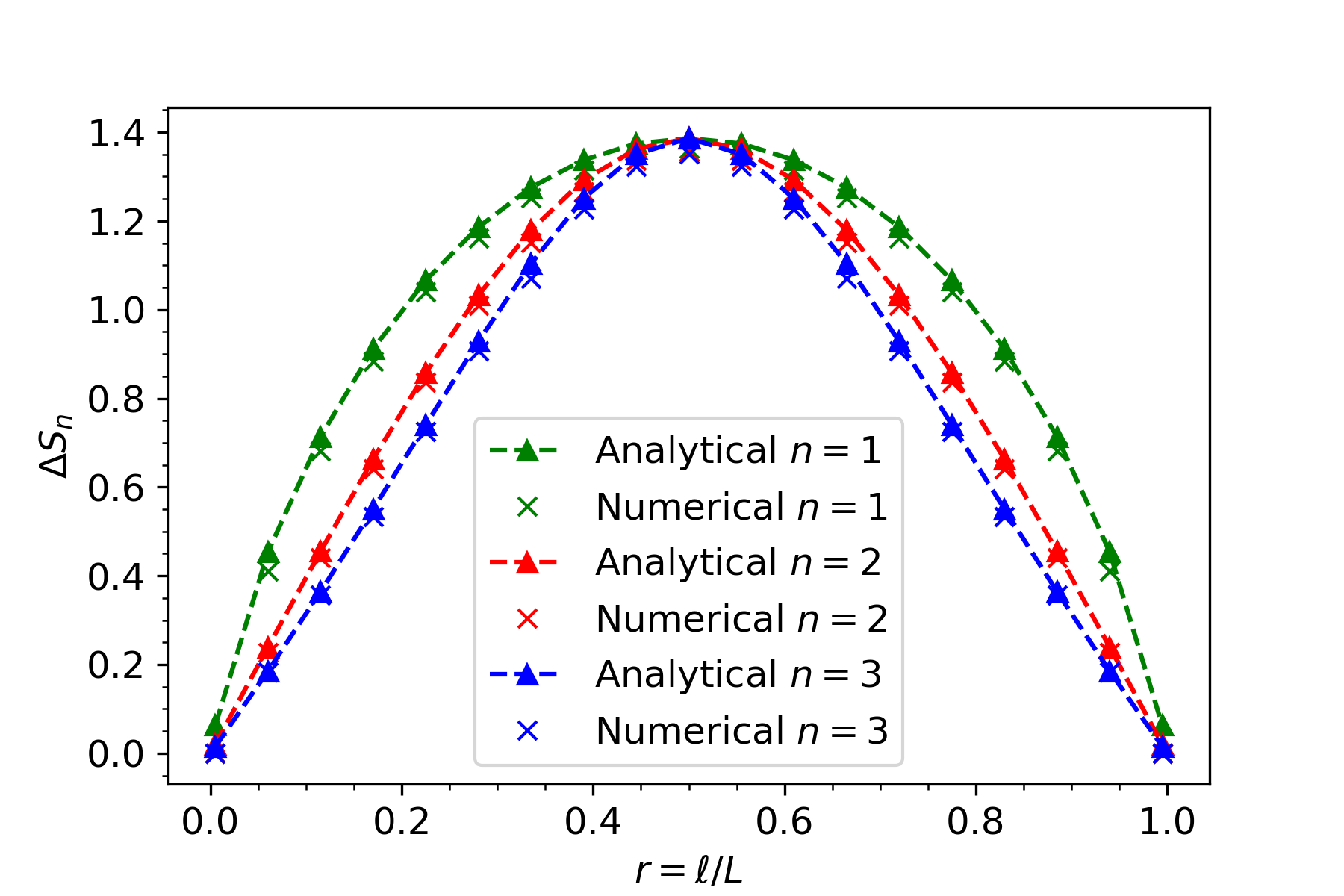}
	\end{subfigure}
	\begin{subfigure}{.5\textwidth}
	\includegraphics[width=\linewidth]{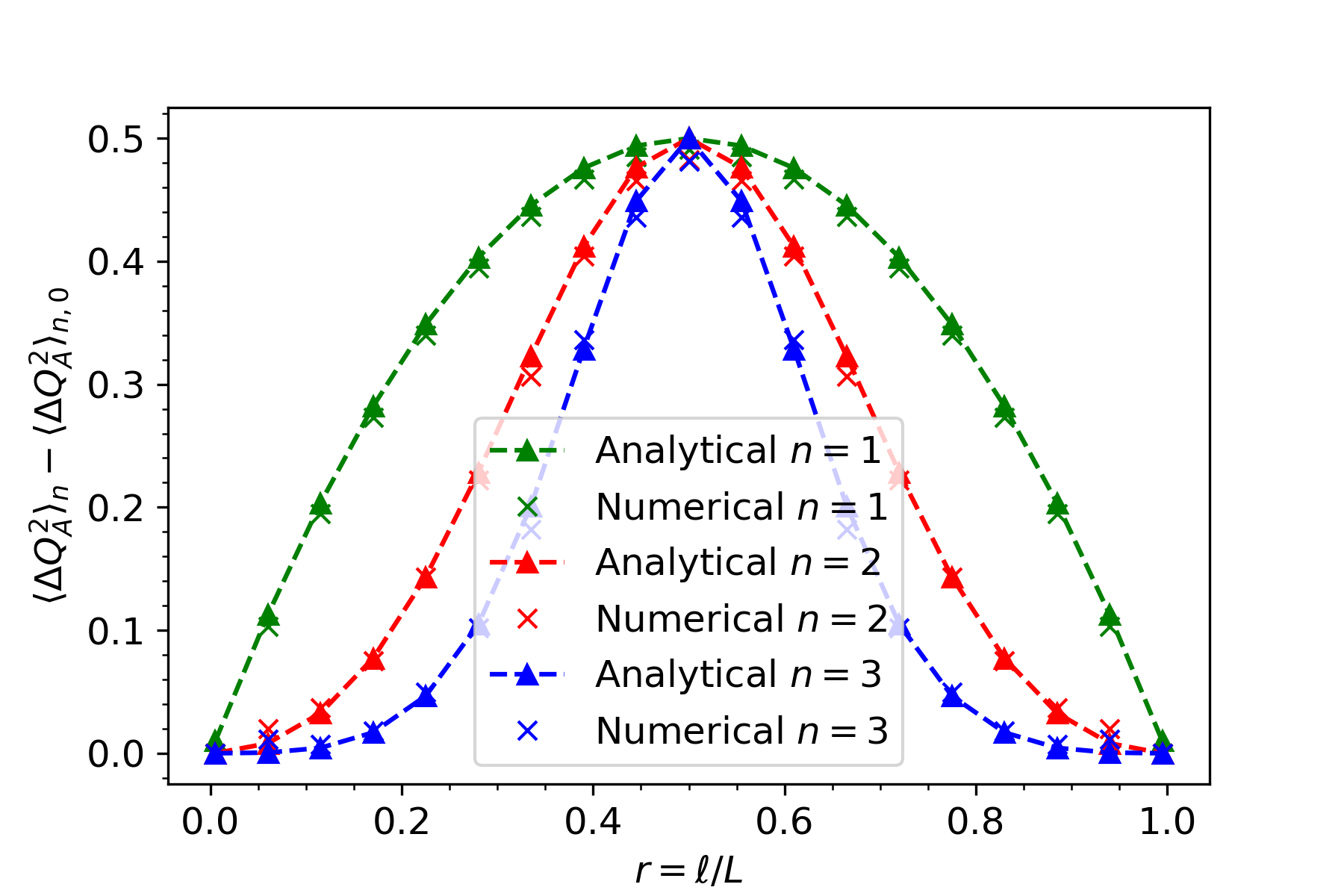}
    \end{subfigure}
    \caption{Numerical data versus analytical prediction for the particle-hole excited state described by the correlation function \eqref{Corr_exc}. The data is for $k_F = \pi/2$, $L=200$ and different values of $n$ for  $r =\ell/L \in [0,1]$. Left: Excess R\'enyi entropy checked against Eq. \eqref{delta_Sn}. Right: Excess variance, checked against Eq. \eqref{Var_exc}. The numerical results are in very good agreement with the analytical formulae.}
    \label{fig_exc}
\end{figure}

\subsection{Complex Free Bosons and the Wave-Functional Method}
\label{waveF}
In this section we consider a complex massive free boson. Unlike the 1D Fermi gas, this model and its lattice version allow us to test formulae for states containing two or more identical excitations. Furthermore, thanks to the computation technique that we outline below, we are able to directly access the ratios
 $M_n^\Psi (r;\alpha)$ defined in \eqref{eq_Mratio} for different values of $\alpha$. Our numerical computation is based on the wave-functional method introduced in \cite{excited2} (see Appendix A of that paper). Here we need to extend the technique to a complex theory and to the symmetry resolved moments. These extensions are not entirely trivial and for that reason we review the wave-functional method in detail. 

\medskip
Let us consider a 1D complex massive boson on the line $[0,L]$ with Hamiltonian:
\be
\label{complex_boson_Hamiltonian}
H = \int_0^L \mathrm{d}x \left(\Pi^\dagger \Pi + (\partial_x \Phi)^\dagger (\partial_x \Phi) + m^2\Phi^\dagger \Phi \right),
\ee
where
\be
\Pi(x) = \dot\Phi^\dagger(x) \quad , \quad \Pi^\dagger(x) = \dot\Phi(x) \,.
\ee
Alternatively, we can introduce a pair of real bosons $\Phi_1,\Phi_2$, and express $\Phi$ and $\Pi$ as
\be
\label{real_dofs}
\Phi = \frac{\Phi_1+i\Phi_2}{\sqrt{2}} \quad , \quad \Pi = \frac{\Pi_1+i\Pi_2}{\sqrt{2}}\,,
\ee
so that the Hamiltonian becomes that of two real bosons. The only non-vanishing equal-time commutators are:
\be
[\Phi(x),\Pi(y)] = [\Phi^\dagger(x), \Pi^\dagger(y)] = i\delta(x-y) \,.
\ee
Since space is compact, there are discrete energy levels with dispersion relation:
\be
E_p = \sqrt{m^2 + p^2} \quad , \quad p \in \frac{2\pi}{L}\mathbb{Z} \,,
\ee
and the Hamiltonian is diagonalised via the introduction of two sets of creation/annihilation operators. The annihilation operators are:
\begin{align}
A_p &= \frac{1}{\sqrt{2LE_p}} \int_0^L \mathrm{d}x e^{-ipx}(E_p \Phi(x) + i \Pi^\dagger(x)) \\
B_p &= \frac{1}{\sqrt{2LE_p}} \int_0^L \mathrm{d}x e^{-ipx}(E_p \Phi^\dagger(x) + i \Pi(x)) 
\end{align}
and as usual $A^\dagger_p$ ($B^\dagger_p$ resp.) creates a positively (negatively) $U(1)$-charged particle  with momentum $p$ from the vacuum. These operators satisfy:
\be
\label{canonical_commutators}
[A_p, A^\dagger_{p'}] = [B_p, B^\dagger_{p'}] = \delta_{p,p'}\,.
\ee
$\Phi(x)$, $\Pi(x)$ then admit the usual Fourier decomposition
\begin{align}
\Phi(x) &= \sum_p \frac{1}{\sqrt{2LE_p}}\left(A_p e^{ipx} + B^\dagger_p e^{-ipx}\right) \\
\Pi(x) &= -i\sum_p \sqrt{\frac{E_p}{2L}}\left(B_p e^{ipx} - A^\dagger_p e^{-ipx}\right)\,.
\end{align}
Finally, the charge operator corresponding to the $U(1)$ symmetry of the theory is:
\begin{align}
Q = i\int_0^L \mathrm{d}x \, : \left(\Phi^\dagger(x)\Pi^\dagger(x) - \Phi(x)\Pi(x)\right):\,=
 \sum_p (A^\dagger_p A_p - B^\dagger_p B_p)\,,
\end{align}
where the normal ordering means that we subtract every non-operatorial term resulting from the commutation relations (i.e. terms involving $\delta$-functions).

\subsubsection{The Wave-Functional Method}
A very useful way to represent the states in the theory is provided by the wave-functional formalism. In this approach, we associate to every state a functional $\boldsymbol{\Psi}$ acting on the space of \textit{classical} field configurations $\phi\,,\phi^\dagger\,:\, [0,L] \to \mathbb{C}$ and formally defined by:
\be
{\boldsymbol{\Psi}}[\phi,\phi^\dagger] = \bra \phi, \phi^\dagger | \boldsymbol{\Psi} \ket \,.
\ee
The action of the operators $\Phi$, $\Pi$ on the (rather abstract) state $| \phi, \phi^\dagger \ket$ mimics the action of the position and momentum operators on the state $|x\ket$ in non-relativistic quantum mechanics, so that in the wave-functional representation:
\be
\Phi(x){\boldsymbol{\Psi}}[\phi, \phi^\dagger] = \phi(x){\boldsymbol{\Psi}}[\phi, \phi^\dagger] \quad , \quad i\Pi(x){\boldsymbol{\Psi}}[\phi, \phi^\dagger] = \frac{\delta {\boldsymbol{\Psi}}[\phi, \phi^\dagger]}{\delta \phi (x)}\,,
\ee
and analogously
\be
\Phi^\dagger(x){\boldsymbol{\Psi}}[\phi, \phi^\dagger] = \phi^\dagger(x){\boldsymbol{\Psi}}[\phi, \phi^\dagger] \quad , \quad i\Pi^\dagger(x){\boldsymbol{\Psi}}[\phi, \phi^\dagger] = \frac{\delta {\boldsymbol{\Psi}}[\phi, \phi^\dagger]}{\delta \phi^\dagger(x)}\,.
\ee
Notice that, for consistency, when $i\Pi$, $i\Pi^\dagger$ act on a ket $|\phi, \phi^\dagger\ket$ there is a minus sign in front of the functional derivative. The functional of the vacuum state is defined by:
\be
A_p {\boldsymbol{\Psi}}_{\text{vac}} = B_p {\boldsymbol{\Psi}}_{\text{vac}}  = 0 \quad \forall \,p\,,
\ee
and the only solution to these functional differential equations up to normalisation is the Gaussian functional:
\be
\label{vacuum_functional}
{\boldsymbol{\Psi}}_{\text{vac}}[\phi,\phi^\dagger] = \exp{\left[-\int_0^L \mathrm{d}x\mathrm{d}y\, \phi^\dagger(x)K(x-y)\phi(y) \right]} \quad , \quad K(x-y)= \frac{1}{L}\sum_p E_p e^{ip(x-y)}\,.
\ee
Notice that $K(x)$ is a real, even function of $x$. The functionals of the positively and negatively charged one-particle states are obtained through the action of $A^\dagger_p$ and $B^\dagger_p$:
\begin{align}
A^\dagger_p {\boldsymbol{\Psi}}_{\text{vac}} &= \alpha_p[\phi^\dagger]{\boldsymbol{\Psi}}_{\text{vac}} \quad , \quad \alpha_p[\phi^\dagger] = \sqrt{\frac{2E_p}{L}}\int_0^L \mathrm{d}x\,e^{ipx} \phi^\dagger(x) \\
B^\dagger_p {\boldsymbol{\Psi}}_{\text{vac}} &= \beta_p[\phi]{\boldsymbol{\Psi}}_{\text{vac}} \quad , \quad \beta_p[\phi] = \sqrt{\frac{2E_p}{L}}\int_0^L \mathrm{d}x\,e^{ipx} \phi(x)\,,
\end{align}
and the functional for a state with $k^+$ positive excitations and $k^-$ negative excitations (all with different momenta) is:
\be
\label{multi-particle state}
{\boldsymbol{\Psi}}^{k^+,k^-}_{p_i,q_j}[\phi, \phi^\dagger] = \prod_{i=1}^{k^+} A^\dagger_{p_i} \prod_{j=1}^{k^-} B^\dagger_{q_j} {\boldsymbol{\Psi}}_\text{vac} = \prod_{i=1}^{k^+} \alpha_{p_i}[\phi^\dagger] \prod_{j=1}^{k^-} \beta_{q_j}[\phi] {\boldsymbol{\Psi}}_\text{vac}[\phi,\phi^\dagger]\,.
\ee
A correct choice of the normalisation in \eqref{vacuum_functional} ensures that the functional above has unit norm with respect to the bra-ket product. However, the normalisation must be modified when some of the particles' momenta are equal, according to \eqref{canonical_commutators}. If there are $k_i^+$ ($k_i^-$) positively (negatively) charged particles with momentum $p_i$ ($q_i$), for $i=1,\dots,m^+$ ($m^-$) we define:
\be
\label{multi-particle state_equal_momenta}
{\boldsymbol{\Psi}}^{k^+,k^-}_{p_i,q_j}[\phi, \phi^\dagger] = \prod_{i=1}^{m^+} \frac{(A^\dagger_{p_i})^{k_i^+}}{\sqrt{k_i^+!}} \prod_{j=1}^{m^-} \frac{(B^\dagger_{q_j})^{k_j^-}}{\sqrt{k_j^-!}} {\boldsymbol{\Psi}}_\text{vac} 
\ee
with $\sum_{i=1}^{m^+}k_i^+=k^+$, $\sum_{i=1}^{m^-}k_i^-=k^-$.
The action of the charge operator on a wave-functional immediately follows from that of the fields and their conjugates:
\be
Q{\boldsymbol{\Psi}}[\phi,\phi^\dagger] = \int_0^L \mathrm{d}x\,\left(\phi^\dagger(x)\frac{\delta}{\delta \phi^\dagger(x)} - \phi(x)\frac{\delta}{\delta \phi(x)}\right){\boldsymbol{\Psi}}[\phi,\phi^\dagger]\,,
\ee
and in particular one finds that the vacuum and the Fock state functionals are charge eigenstates:
\be
Q{\boldsymbol{\Psi}}_\text{vac}=0 \quad, \quad Q{\boldsymbol{\Psi}}^{k^+,k^-}_{p_i,q_j} = (k^+ - k^-){\boldsymbol{\Psi}}^{k^+,k^-}_{p_i,q_j}\,.
\ee
However, because $|\phi, \phi^\dagger\ket$ is not associated to any charged state in the Fock space, it is not a charge eigenstate. The exponential of the charge operator acts on $|\phi, \phi^\dagger\ket$ by introducing phases (notice the minus sign in front of the integral):
\be
\label{Q_action_ket}
e^{2\pi i \alpha Q}|\phi, \phi^\dagger \ket = \exp\left[-2\pi i \alpha \int_0^L \mathrm{d}x\left(\phi^\dagger(x)\frac{\delta}{\delta \phi^\dagger(x)}-\phi(x)\frac{\delta}{\delta \phi(x)}\right)\right]|\phi, \phi^\dagger \ket = |e^{2\pi i \alpha}\phi , e^{-2\pi i \alpha}\phi^\dagger \ket \,.
\ee
This can be argued, for instance, by looking at the corresponding infinitesimal translation:
\be
|\phi + 2\pi i \alpha\, \phi \ket \simeq |\phi\ket + 2\pi i \alpha \int_0^L \mathrm{d}x\,\phi(x)\frac{\delta}{\delta \phi(x)}|\phi \ket\,.
\ee
Employing these results it is possible to show (see Appendix A for the derivation) that 
\be
\Tr_A(\rho^n_{0,A}e^{2\pi i \alpha Q_A}) = \int \mathcal{D}\phi_1\dots \mathcal{D}\phi_n \exp \left[-\mathcal{G}_\alpha\right]\,,
\ee
where $\mathcal{G}_\alpha$ is a known gaussian functional of the fields $\phi(x), \phi^\dagger(x)$ given in (\ref{continuous_matrix_alpha}). Results for the harmonic chain can then be obtained by discretisation, as we see in the next subsection.

\subsection{The Harmonic Chain}
\label{23}
Since the Hamiltonian \eqref{complex_boson_Hamiltonian} reduces to the sum of two Hamiltonians for the real bosons $\Phi_1$, $\Phi_2$ with prefactors $\frac{1}{2}$, the discretisation proceeds exactly as for the real boson \cite{excited2}. We divide the interval $[0,L]$ in $N$ parts introducing a spacing:
\be
\Delta x = \frac{L}{N}
\ee
and we define $x= \frac{L}{N}\bar{x}$, $\bar{x}\,\in\, \{0,1,\dots,N-1\}$, so that we can replace every integral with a sum:
\be
\label{integrals_to_sums}
\int_{A \bigcup {\bar{A}} }\mathrm{d}x \,\rightarrow\, \frac{L}{N}\sum_{x=0}^{L-\Delta x} \quad , \quad 
\int_{A}\mathrm{d}x \,\rightarrow\, \frac{L}{N}\sum_{x=0}^{\ell-\Delta x}\quad , \quad 
\int_{{\bar{A}} }\mathrm{d}x \,\rightarrow\, \frac{L}{N}\sum_{x=\ell}^{L-\Delta x}\,.
\ee
If we discretise the Laplace operator as:
\be
\partial_x^2\Phi(x) \,\rightarrow \, \frac{\Phi(x + \Delta x) + \Phi(x-\Delta x) - 2\Phi (x)}{(\Delta x)^2}
\ee
and impose periodic boundary conditions $\Phi(0)=\Phi(L)$, $\Phi^\dagger(0)=\Phi^\dagger(L)$, the Hamiltonian \eqref{complex_boson_Hamiltonian} reduces (upon integration by parts) to two independent harmonic chains for real fields. The set of momenta is now restricted to the first Brillouin zone, $p= \frac{2\pi}{L}\bar{p}\,,\bar{p}\, \in \, \{0,1,\dots, N-1\}$, and the dispersion relation becomes:
\be
E_p = \sqrt{m^2 + \left(\frac{2N}{L}\sin\frac{pL}{2N}\right)^2}\,,
\ee
from which the relativistic relation $E_p^2=m^2+p^2$ is obtained when  $\frac{p L}{2N}\ll 1$.
Notice that since we restrict the set of momenta, the function $K(x)$ defined in \eqref{vacuum_functional} becomes a finite sum:
\be
K(x) = \frac{1}{L}\sum_{p=0}^{2\pi (N-1)/L}E_p e^{ipx}\,,
\ee
thus it is no longer an even function of $x$, though the property $K^*(x)= K(-x)$ still holds. 

For the sake of simplicity we will take $\Phi$ and $\Phi^\dagger$ to be the fundamental degrees of freedom in the following, while keeping in mind that the real degrees of freedom are recovered using \eqref{real_dofs}. In the formula \eqref{ratio_of_charged_moment_functionals} the functions $U_i^\pm$, $V_i^\pm$ are modified by simply replacing the integrals with sums following the prescription \eqref{integrals_to_sums}. On the other hand, discretisation of the measure $\mathcal{G}_\alpha$ leads to a finite-dimensional $(nN)\times (nN)$ matrix $G$ which couples the fields $\phi_i^\dagger(x)$ and $\phi_j(y)$:
\begin{align}
\label{discretized_matrix}
\mathcal{G}_\alpha &= \left(\frac{L}{N}\right)^2\sum_{i=1}^n \left[2\left(\sum_{x \, \in \, A \,, y \, \in \, A} + \sum_{x \, \in \, {\bar{A}}  \,, y \, \in \, {\bar{A}} }\right)\phi_i^\dagger(x)K(x-y)\phi_i(y)  \right. \nonumber \\&+ \left. \sum_{x \, \in \, A \,, y \, \in \, {\bar{A}} } \left(\phi^\dagger_i(x) + \phi^\dagger_{i+1}(x) e^{-2\pi i \alpha \delta_{i,n}}\right) K(x-y)\phi_i(y)  \right. \nonumber \\&+ \left. \sum_{x \, \in \, {\bar{A}}  \,, y \, \in \, A} \phi_i^\dagger(x)K(x-y)\left(\phi_i(y)+\phi_{i+1}(y)e^{2\pi i\alpha \delta_{i,n}}\right)\right] \equiv \sum_{i,j=1}^n\sum_{x,y=0}^L\phi_i^\dagger(x)G_{ix,jy}\phi_j(y)\,,
\end{align}
where we explicitly wrote the complex conjugate in \eqref{continuous_matrix_alpha} before discretising the integrals. Wick's theorem ensures that the Gaussian average in \eqref{ratio_of_charged_moment_functionals} can be computed from the contractions of pairs of fields, which are in turn obtained via the inversion of the matrix $G$:
\be
\label{wick_contraction}
\wick{\c{\phi_i(x)}^\dagger \c{\phi_j(y)}}= (G^{-1})_{ix,jy}\,.
\ee
The matrix $G$ has a block structure, consisting of $n^2$ blocks $G_{i\cdot,j\cdot}$, each of which is an $N \times N$ matrix. From the above expression we see that the only non-vanishing blocks are either in the diagonal $G_{i\cdot,i\cdot}$ or just off the diagonal, $G_{i\cdot,(i\pm 1)\cdot}$. Each block G admits a sub-block structure in terms of the matrices  $K_{Q_1Q_2}$, $Q_1, Q_2 \,\in\, \{A,{\bar{A}} \}$, whose elements are:
\be
(K_{Q_1Q_2})_{xy} = \left(\frac{L}{N}\right)^2K(x-y) \quad , \quad x \,\in\,Q_1 \,, y \,\in\,Q_2\,.
\ee
Notice that $K_{AA}$ is a square matrix with dimensions $\frac{\ell}{L}N \times \frac{\ell}{L}N$, $K_{A{\bar{A}}}$ has dimensions $\frac{\ell}{L}N \times \frac{L-\ell}{L}N$ and so on. From \eqref{discretized_matrix} we obtain the following basic structures:
\begin{itemize}
    \item Diagonal blocks $G_{i\cdot,i\cdot}$ \\
    \begin{align*}
    &\left(\sum_{x \, \in \, A \,, y \, \in \, A} + \sum_{x \, \in \, \bar{A} \,, y \, \in \, \bar{A}}+ \sum_{x \, \in \, A \,, y \, \in \, \bar{A}}+ \sum_{x \, \in \, \bar{A} \,, y \, \in \, A}\right)\phi_i^\dagger(x)G_{ix,iy}\phi_i(y) \\ 
    &=\left(\frac{L}{N}\right)^2\left[ 2 \sum_{\substack{x \, \in \, A \\ y \, \in \, A}}\phi_i^\dagger(x)K(x-y)\phi_i(y) + 2 \sum_{\substack{x \, \in \, \bar{A}\\ y \, \in \, \bar{A}}}\phi_i^\dagger(x)K(x-y)\phi_i(y) \right. \\ &+\left.  \sum_{x \, \in \, A \,, y \, \in \, \bar{A}}\phi_i^\dagger(x)K(x-y)\phi_i(y) + \sum_{x \, \in \, \bar{A} \,, y \, \in \, A}\phi_i^\dagger(x)K(x-y)\phi_i(y)\right]\\
    &\Rightarrow G_{i\cdot,i\cdot} = 
    \begin{pmatrix}
    2K_{AA} & K_{A{\bar{A}}}  \\
     K_{\bar{A} A} & 2K_{\bar{A} \bar{A}}
    \end{pmatrix}
    \end{align*}
    \item Off-diagonal blocks $G_{i\cdot,(i+1)\cdot}$ \\
    \begin{align*}
    & \sum_{x \, \in \, \bar{A} \,, y \, \in \, A}\phi_i^\dagger(x)G_{ix,(i+1)y}\phi_{i+1}(y) =\left(\frac{L}{N}\right)^2 \sum_{x \, \in \, \bar{A} \,, y \, \in \, A}\phi_i^\dagger(x)K(x-y)\phi_{i+1}(y)e^{2\pi i \alpha \delta_{i,n}}\\
    &\Rightarrow G_{i\cdot,(i+1)\cdot} = 
    \begin{pmatrix}
     0 & 0  \\
     K_{\bar{A}A}e^{2\pi i \alpha \delta_{i,n}} & 0
    \end{pmatrix}
    \end{align*}
    \item Off-diagonal blocks $G_{(i+1)\cdot,i\cdot}$ \\
    \begin{align*}
    & \sum_{x \, \in \, A \,, y \, \in \, \bar{A}}\phi_{i+1}^\dagger(x)G_{(i+1)x,iy}\phi_i(y) =\left(\frac{L}{N}\right)^2 \sum_{x \, \in \, A \,, y \, \in \, \bar{A}}\phi_{i+1}^\dagger(x)K(x-y)\phi_i(y)e^{-2\pi i \alpha \delta_{i,n}}\\
    &\Rightarrow G_{(i+1)\cdot,i\cdot} = 
    \begin{pmatrix}
     0 & K_{A\bar{A}}e^{-2\pi i \alpha \delta_{i,n}}  \\
     0 & 0
    \end{pmatrix}
    \end{align*}
\end{itemize}
Note that the block structure is different from that in \cite{excited2} because the roles of regions $A$ and $\bar{A}$ are now exchanged. Although this exchange has no effect on the form of the entanglement entropy, for the SREE it makes a difference as the symmetry between  $A, \bar{A}$ is broken when we choose to place the charge in subsystem $A$. 

In terms of the $N \times N$ blocks above, we can schematically write the matrix $G$ as follows:

\renewcommand{\arraystretch}{1.4}
\begin{tabular}{ r r V{2} c : c V{2} c : c V{2} c V{2} c : c V{2} c : c V{2}}
\multicolumn{2}{c}{} & \multicolumn{2}{c}{$1$} & \multicolumn{2}{c}{$2$} & \multicolumn{1}{c}{}& \multicolumn{2}{c}{$n-1$} & \multicolumn{2}{c}{$n$} 
\\[-1.2em] 
\multicolumn{2}{c}{} & \multicolumn{2}{c}{\downbracefill} & \multicolumn{2}{c}{\downbracefill} & \multicolumn{1}{c}{}& \multicolumn{2}{c}{\downbracefill} & \multicolumn{2}{c}{\downbracefill} 
\\[-0.5em] 
\multicolumn{1}{c}{} &\multicolumn{1}{c}{} & \multicolumn{1}{c}{$A$} & \multicolumn{1}{c}{$\bar{A}$} & \multicolumn{1}{c}{$A$} & \multicolumn{1}{c}{$\bar{A}$} & \multicolumn{1}{c}{} 
& \multicolumn{1}{c}{$A$} & \multicolumn{1}{c}{$\bar{A}$}& \multicolumn{1}{c}{$A$} & \multicolumn{1}{c}{$\bar{A}$}
\\[-1.2em] 
\multicolumn{1}{c}{} &\multicolumn{1}{c}{} & \multicolumn{1}{c}{\downbracefill} & \multicolumn{1}{c}{\downbracefill} & \multicolumn{1}{c}{\downbracefill} & \multicolumn{1}{c}{\downbracefill} & \multicolumn{1}{c}{} & \multicolumn{1}{c}{\downbracefill} & \multicolumn{1}{c}{\downbracefill} & \multicolumn{1}{c}{\downbracefill} & \multicolumn{1}{c}{\downbracefill} 
\\ \clineB{3-6}{2} \clineB{8-11}{2} 
\ldelim\{{2}{0.5em}[$1$] & \ldelim\{{1}{1.1em}[$A$]& $2 K_{AA}$ & $K_{A\bar{A}}$ & $0$ & $0$ & \multirow{2}{*}{ $\cdots$} & $0$ & $0$ & $0$ & $\lambda^*_{\alpha} K_{A\bar{A}}$
\\ \cdashline{3-6} \cdashline{8-11}
& \ldelim\{{1}{1.1em}[$\bar{A}$] & $K_{\bar{A}A}$ & $2 K_{\bar{A}\bar{A}}$ & $K_{\bar{A}{A}}$ & $0$ & & $0$ & $0$ & $0$ & $0$
\\ \clineB{3-6}{2} \clineB{8-11}{2} 
\ldelim\{{2}{0.5em}[$2$] &\ldelim\{{1}{1.1em}[$A$] & $0$ & $K_{A\bar{A}}$ & $2 K_{AA}$ & $K_{A\bar{A}}$ & \multirow{2}{*}{ $\cdots$} & $0$ & $0$ & $0$ & $0$
\\ \cdashline{3-6} \cdashline{8-11}
& \ldelim\{{1}{1.1em}[$\bar{A}$] & $0$ & $0$ & $K_{\bar{A}A}$ & $2 K_{\bar{A}\bar{A}}$ & & $0$ & $0$ & $0$ & $0$
\\ \clineB{3-6}{2} \clineB{8-11}{2} 
\multicolumn{2}{c}{} & \multicolumn{2}{c}{$\vdots$} & \multicolumn{2}{c}{$\vdots$} & \multicolumn{1}{c}{$\ddots$}& \multicolumn{2}{c}{$\vdots$} & \multicolumn{2}{c}{$\vdots$} 
\\ \clineB{3-6}{2} \clineB{8-11}{2} 
\ldelim\{{2}{2.6em}[$n-1$] &\ldelim\{{1}{1.1em}[$A$] & $0$ & $0$ & $0$ & $0$ &\multirow{2}{*}{ $\cdots$} & $2 K_{AA}$ & $ K_{A\bar{A}}$ & $0$ & $0$ 
\\ \cdashline{3-6} \cdashline{8-11}
& \ldelim\{{1}{1.1em}[$\bar{A}$] & $0$ & $0$ & $0$ & $0$ & & $K_{\bar{A}A}$& $2 K_{\bar{A}\bar{A}}$ & $K_{\bar{A}A}$ & $0$
\\ \clineB{3-6}{2} \clineB{8-11}{2} 
\ldelim\{{2}{0.9em}[$n$] &\ldelim\{{1}{1.1em}[$A$] & $0$ & $0$ & $0$ & $0$ &\multirow{2}{*}{ $\cdots$} & $0$ & $K_{A\bar{A}}$ & $2 K_{AA}$ & $K_{A\bar{A}}$
\\ \cdashline{3-6} \cdashline{8-11}
& \ldelim\{{1}{1.1em}[$\bar{A}$] & $\lambda_{\alpha}K_{\bar{A}A}$ & $0$ & $0$ & $0$ & & $0$ & $0$ & $K_{\bar{A}A}$ & $2 K_{\bar{A}\bar{A}}$
\\ \clineB{3-6}{2} \clineB{8-11}{2} \\[-10 pt]
\end{tabular}
\renewcommand{\arraystretch}{1}\\
where we introduced $\lambda_{\alpha}=e^{2\pi i \alpha}$.

\begin{figure}[h!]
	\begin{subfigure}{.5\textwidth}
	\includegraphics[width=\linewidth]{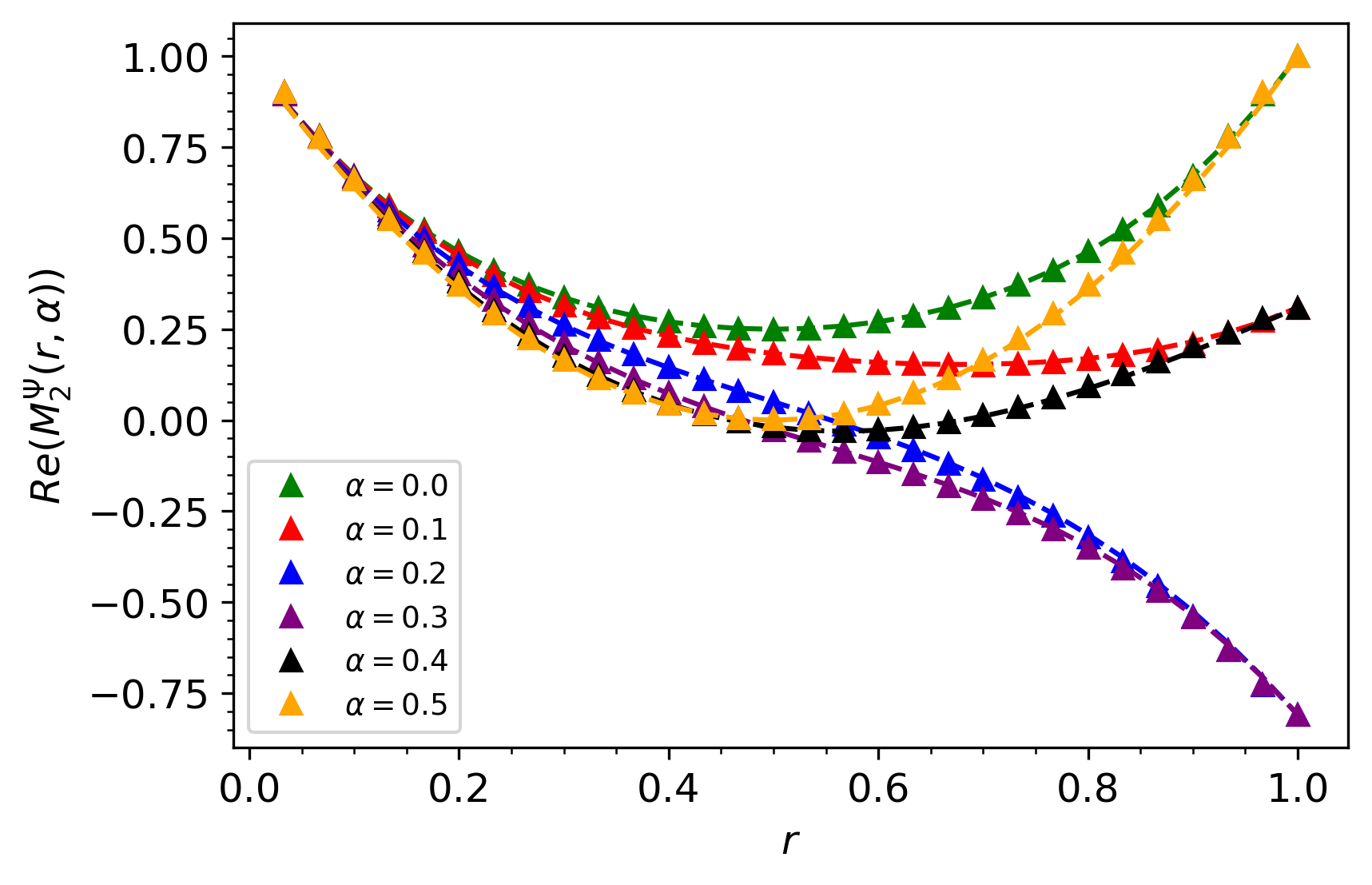}
	\end{subfigure}
	\begin{subfigure}{.5\textwidth}
	\includegraphics[width=\linewidth]{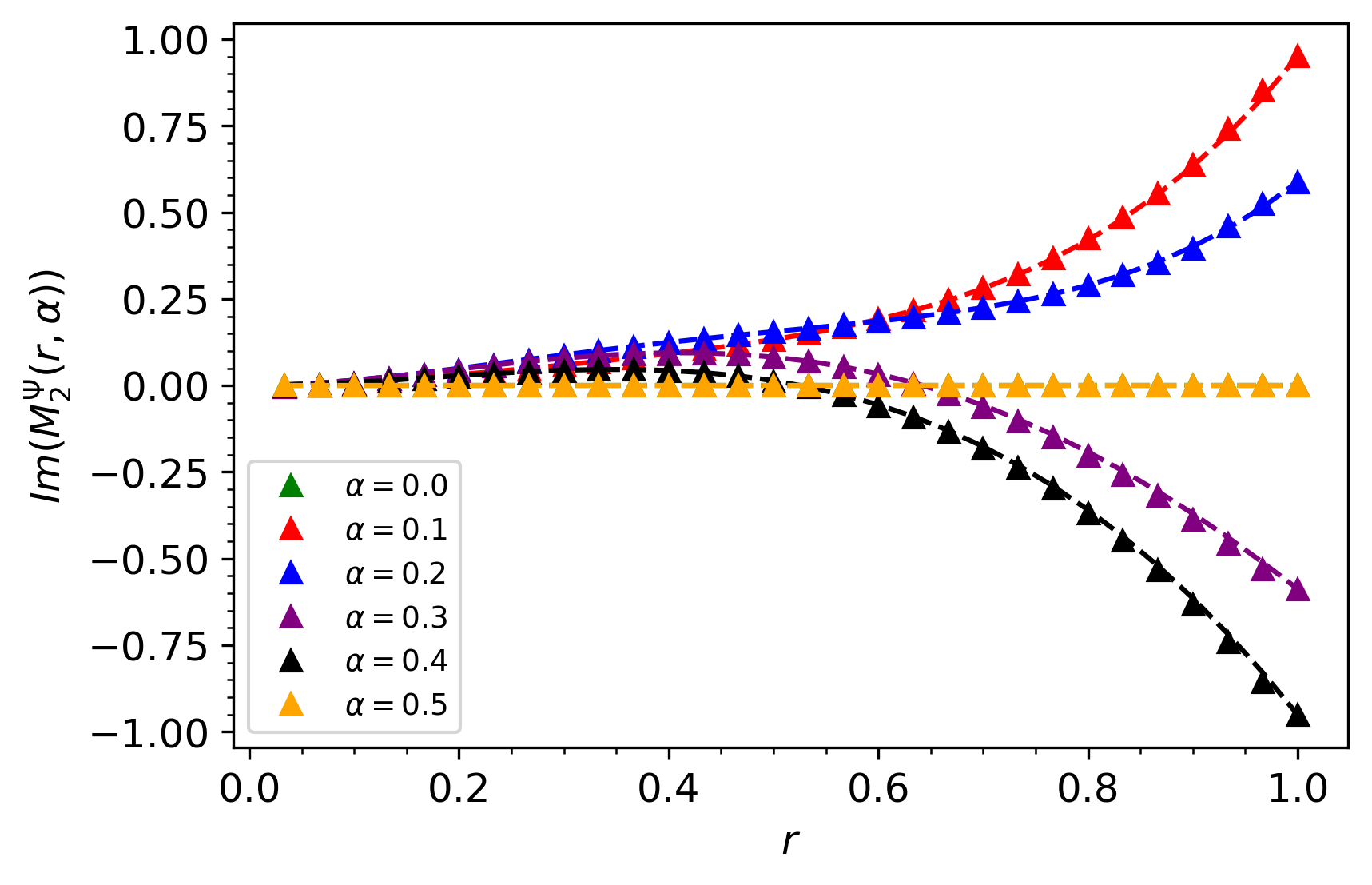}
    \end{subfigure}
    \begin{subfigure}{.5\textwidth}
	\includegraphics[width=\linewidth]{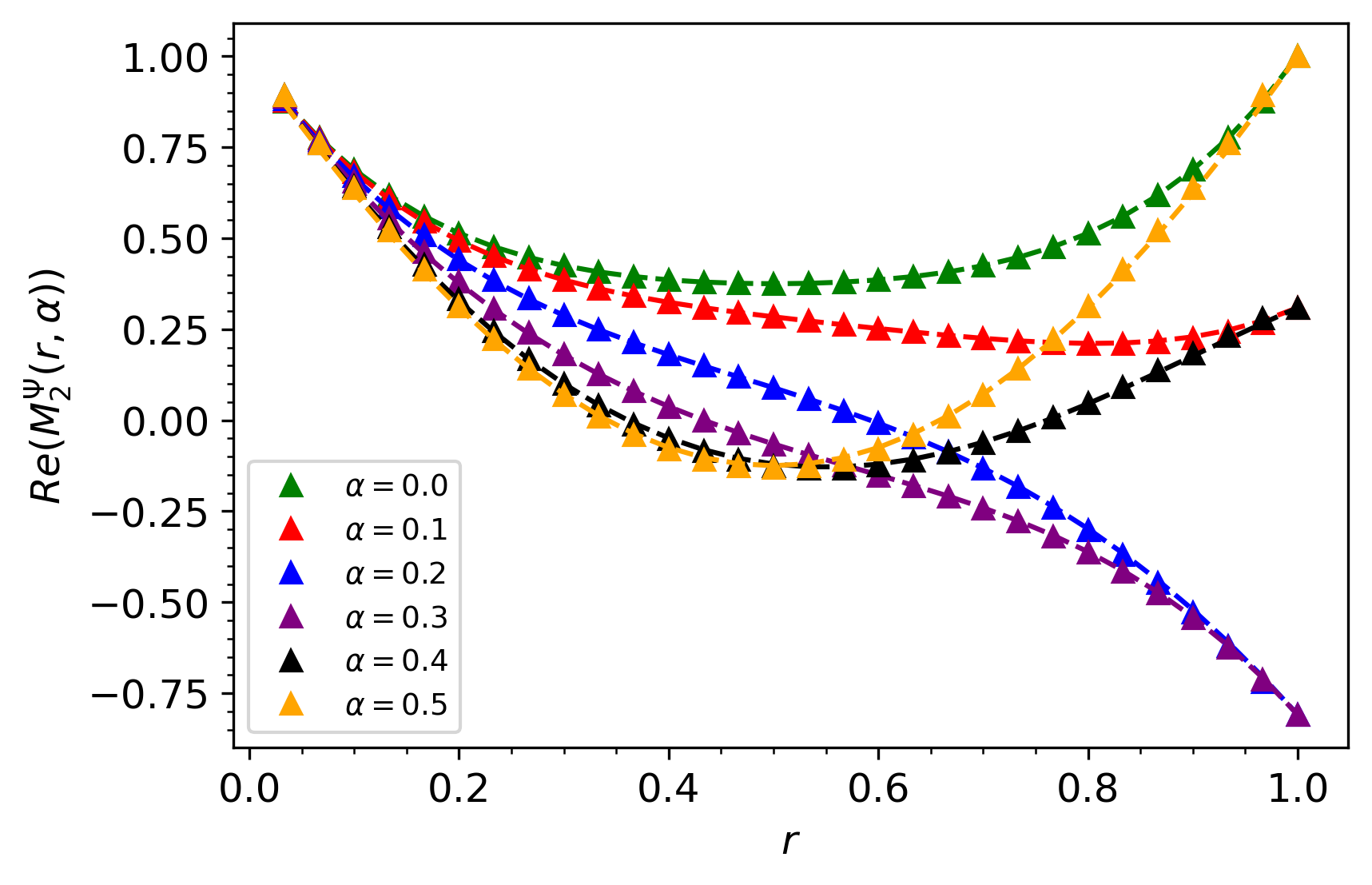}
	\end{subfigure}
	\begin{subfigure}{.5\textwidth}
	\includegraphics[width=\linewidth]{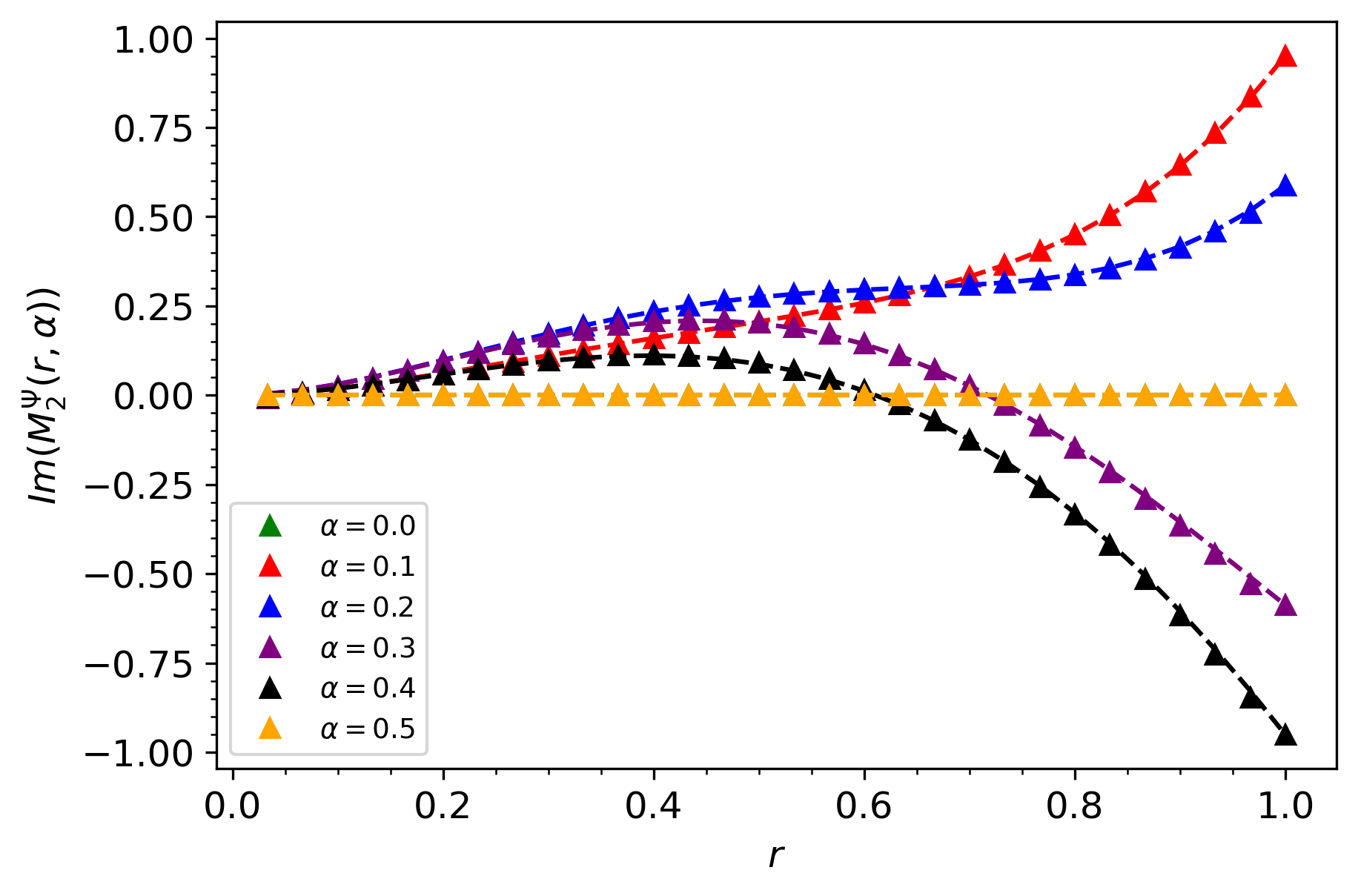}
    \end{subfigure}
    \caption{Numerical data (triangles) versus analytical predictions (dashed lines) for  $M_2^{1^+1^+}(r;\alpha)$ (top row) and $M_2^{2^+}(r;\alpha)$  (bottom row). We consider $n=2$, system size $L = 30$ with $m=0.1$.  The left/right panels in each row show the real/imaginary part of the function. In both rows we take values of the flux  $\alpha=0,0.1,\dots 0.5$. The numerics for the top row figures employ momenta $p_1 = \pi,p_2=2\pi/5$ whereas for the bottom row we took equal momenta $p_1=p_2=\pi$.}
    \label{fig_Harmonic2}
\end{figure}

\begin{figure}[h!]
	\begin{subfigure}{.5\textwidth}
	\includegraphics[width=\linewidth]{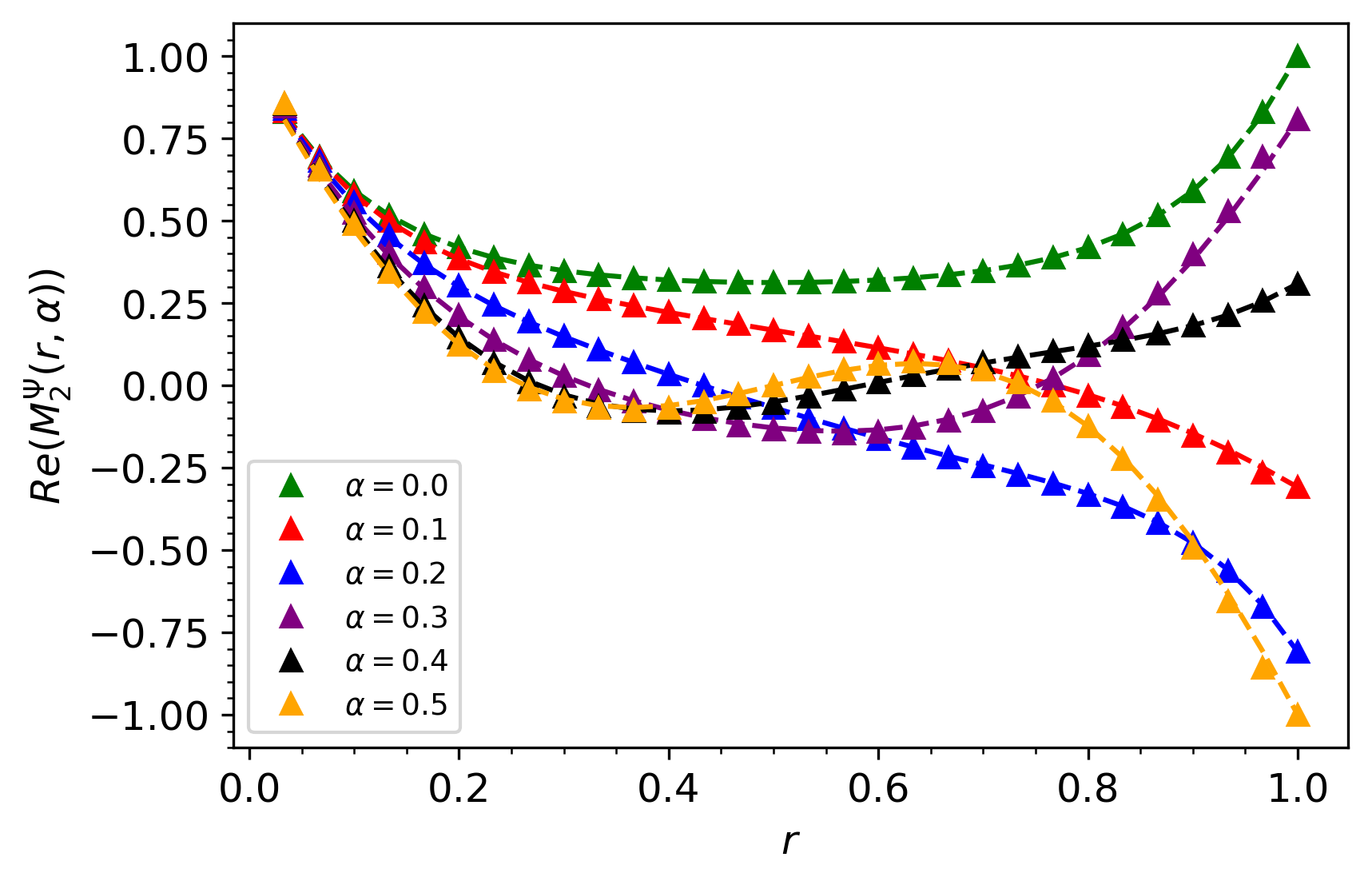}
	\end{subfigure}
	\begin{subfigure}{.5\textwidth}
	\includegraphics[width=\linewidth]{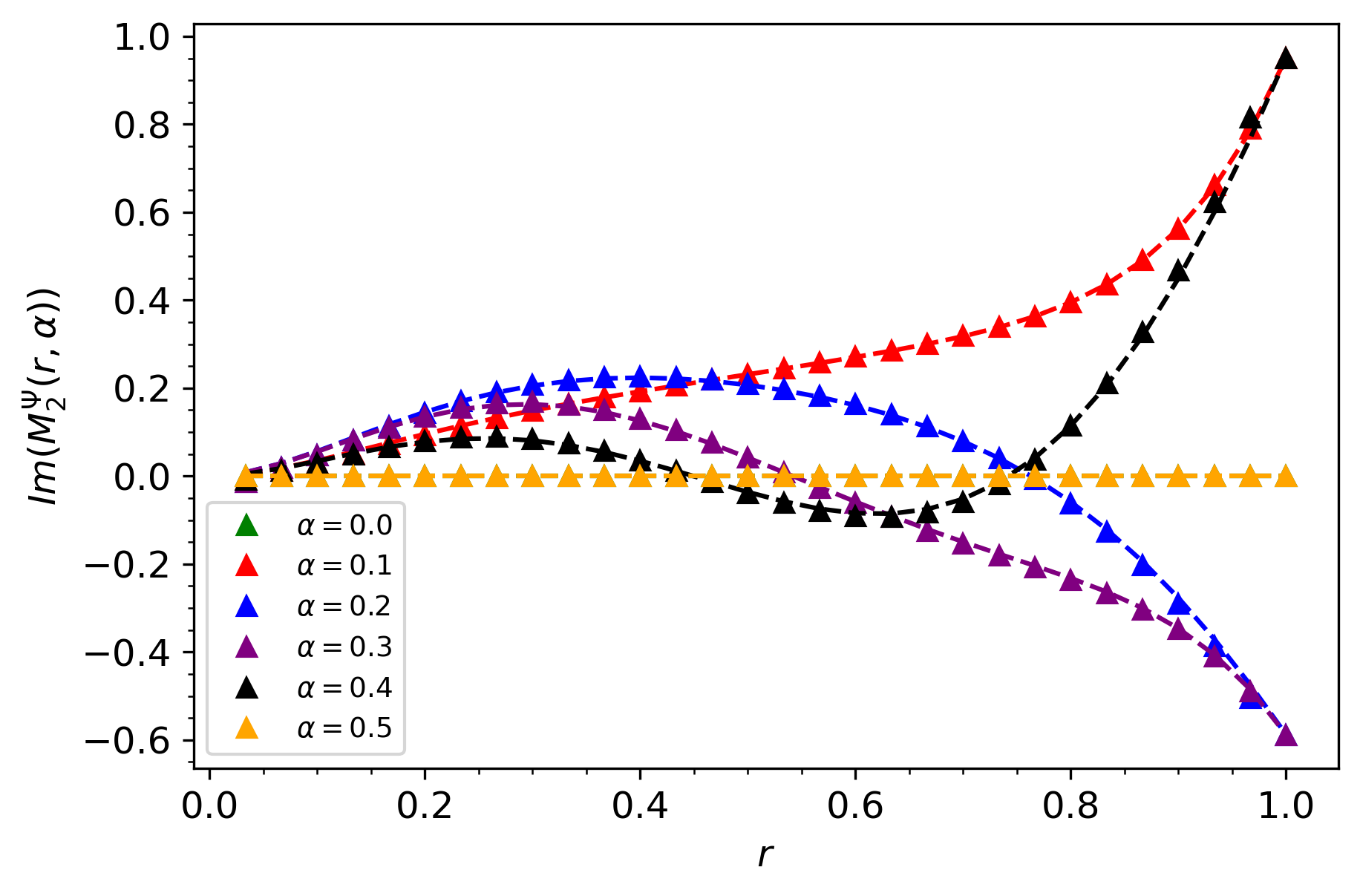}
    \end{subfigure}
    \begin{subfigure}{.5\textwidth}
	\includegraphics[width=\linewidth]{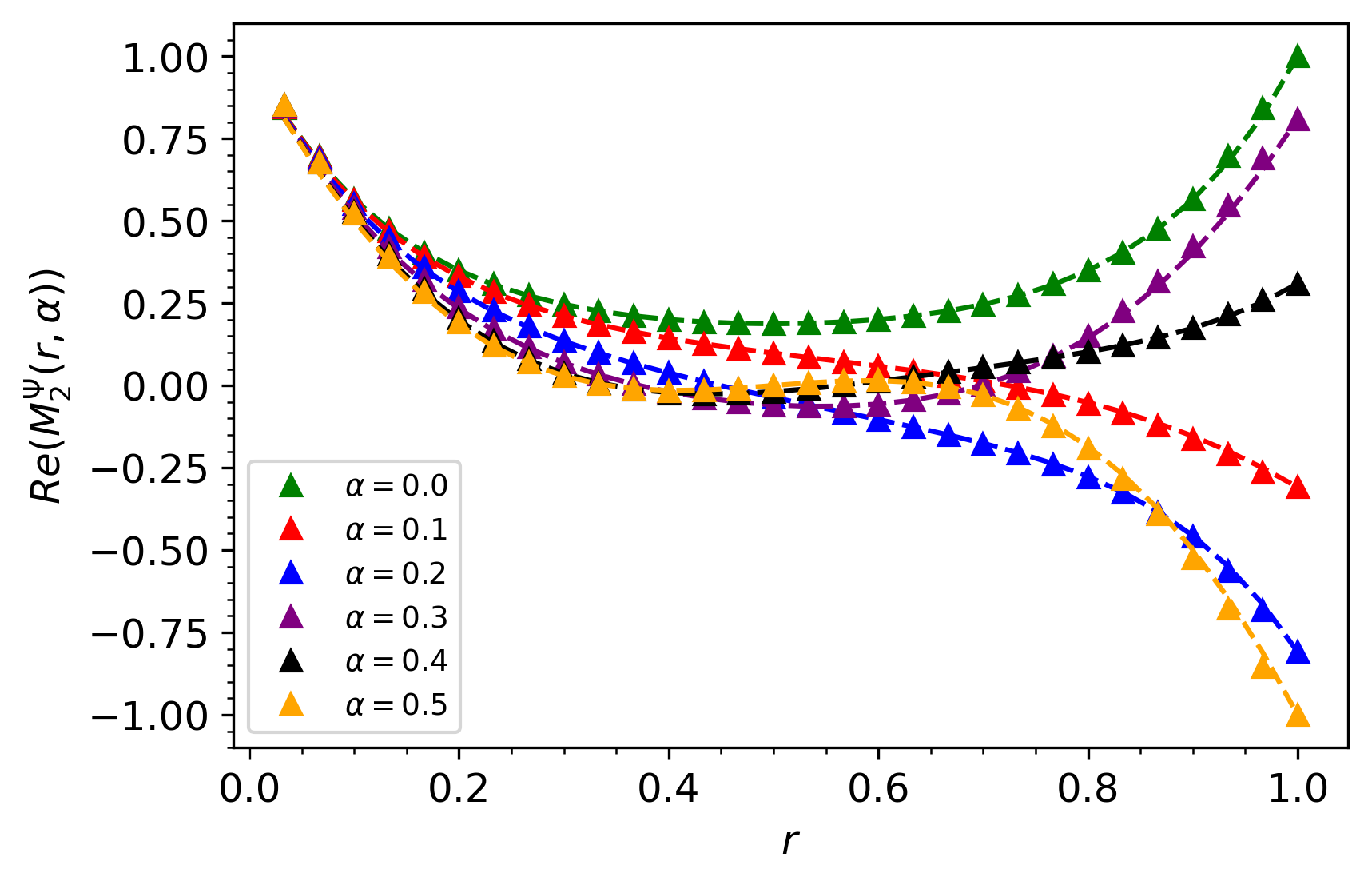}
	\end{subfigure}
	\begin{subfigure}{.5\textwidth}
	\includegraphics[width=\linewidth]{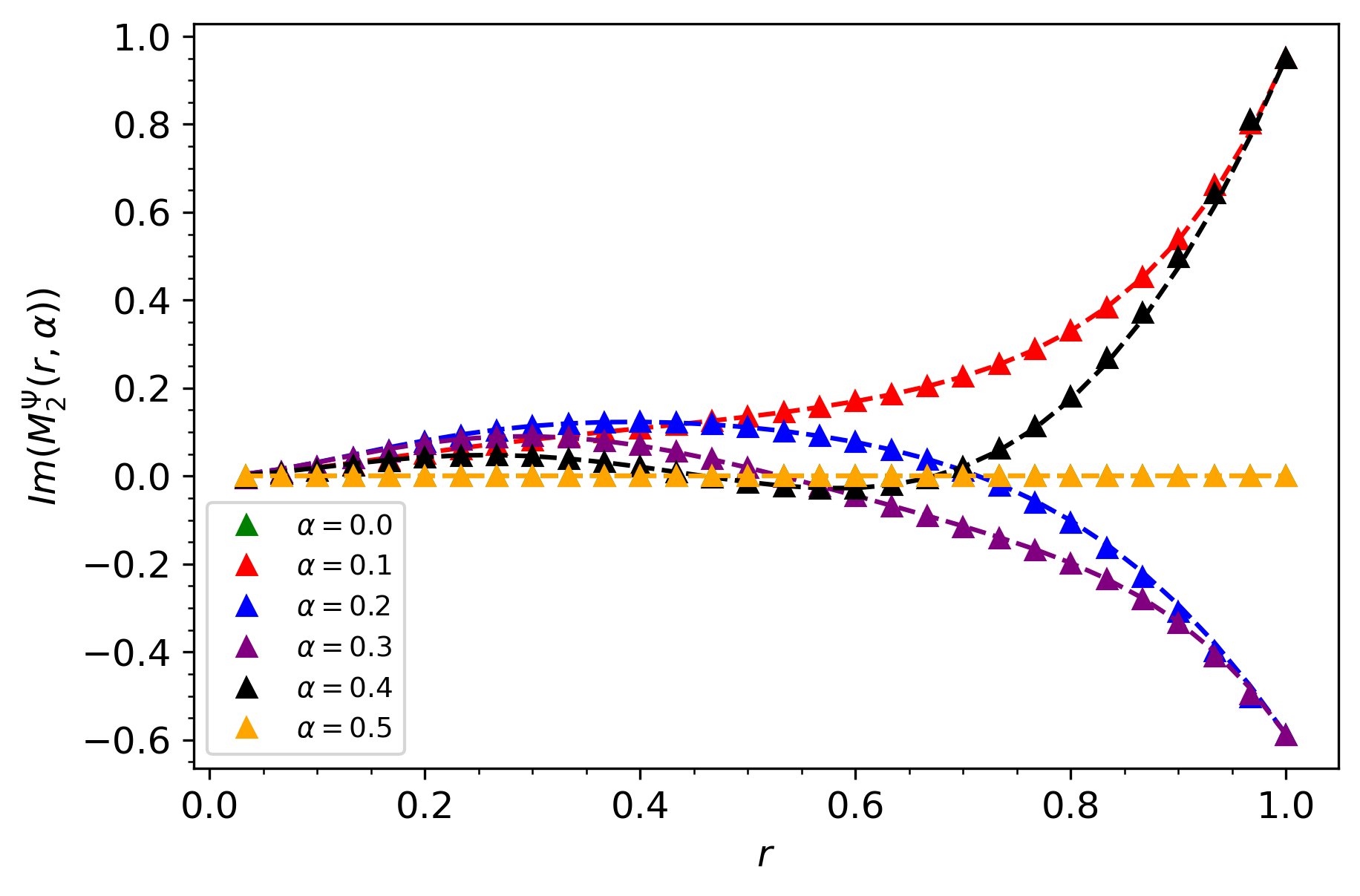}
    \end{subfigure}
    \begin{subfigure}{.5\textwidth}
	\includegraphics[width=\linewidth]{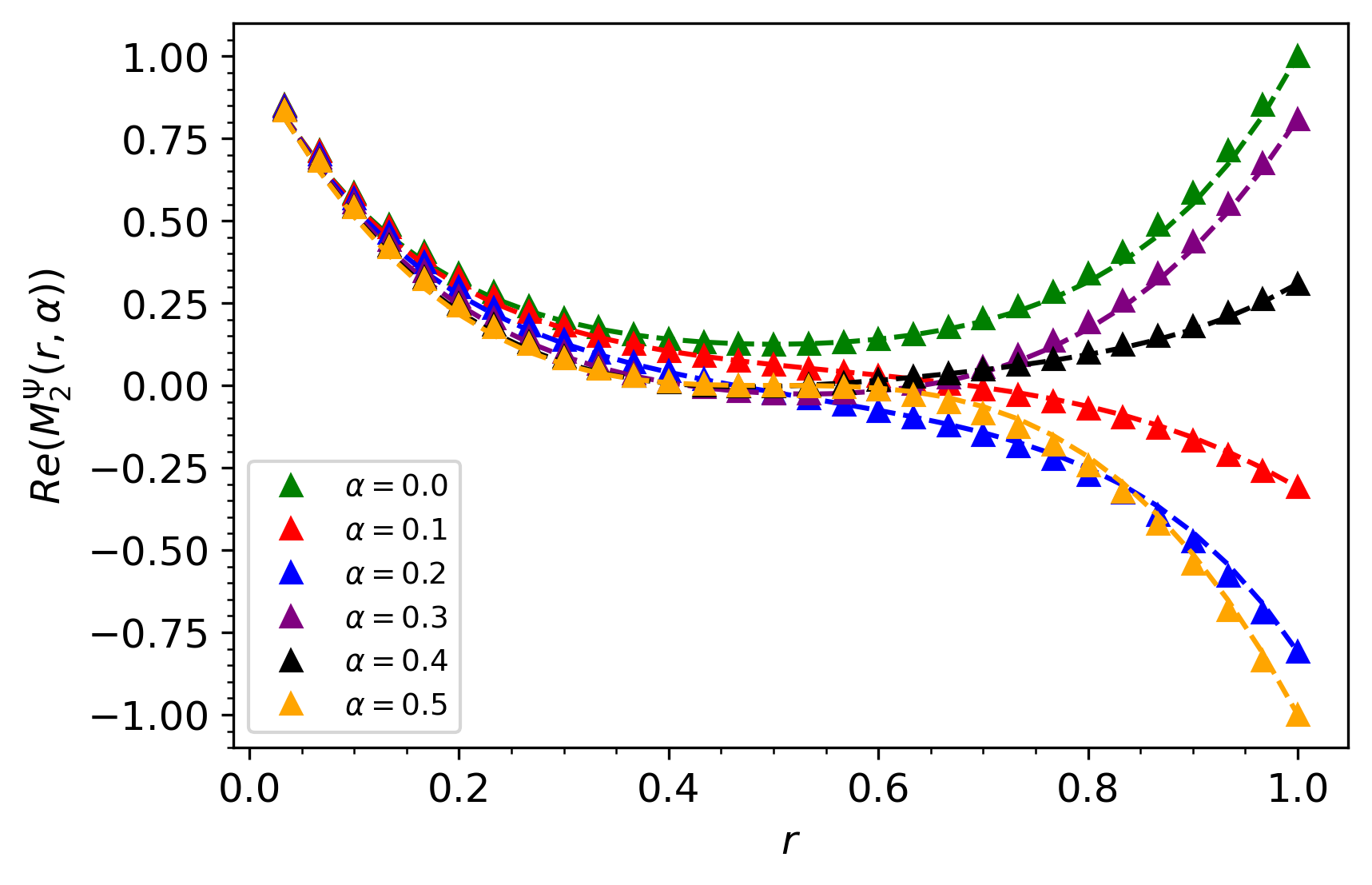}
    \end{subfigure}
    \begin{subfigure}{.5\textwidth}
	\includegraphics[width=\linewidth]{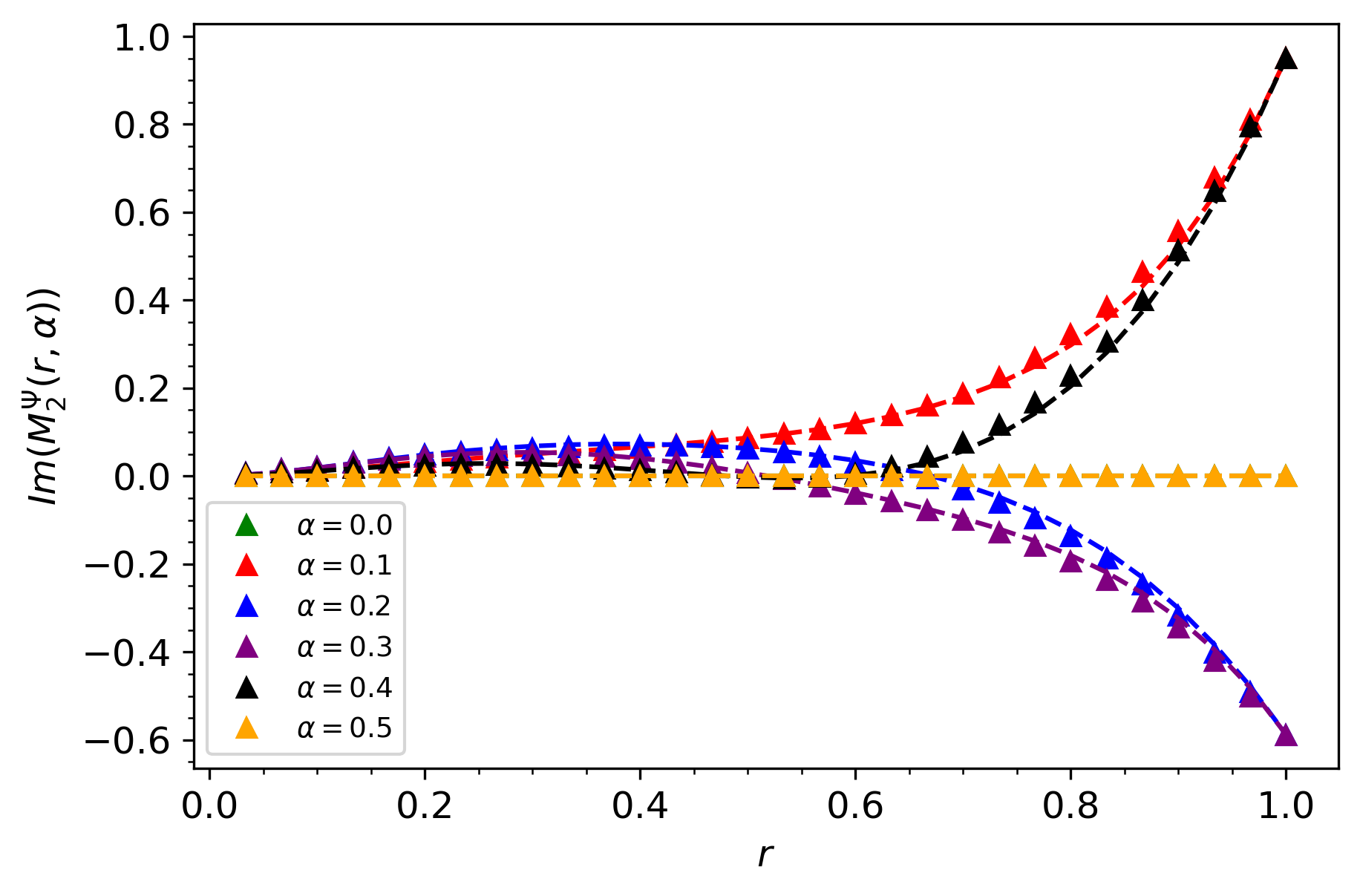}
    \end{subfigure}
    \caption{Numerical data (triangles) versus analytical predictions (dashed lines) for  $M_2^{3^+}(r;\alpha)$ (top row), $M_2^{2^+ 1+}(r;\alpha)$  (central row), and $M_2^{1^+ 1^+ 1^+ }(r;\alpha)$  (bottom row). We consider $n=2$, system size $L = 30$ with $m=0.1$.  The left/right panels in each row show the real/imaginary part of the function. In each rows we take values of the flux  $\alpha=0,0.1,\dots 0.5$. The numerics for the top row figures employ momenta $p_1 = p_2 = p_3 = \pi$, for the central row $p_1 = p_2 = \pi, p_3 = \pi/3$, whereas for the bottom row we took $p_1=\pi, p_2=\pi/3, p_3 =\pi/5$.}
    \label{fig_Harmonic3}
\end{figure}

\subsubsection{Numerical Results}
In Appendix \ref{AppA} we have explicitly derived the ratio of charged moments for excited states. We thus have 
all the ingredients needed to obtain numerical results. 
Let us take a bi-partition where $A$ is a segment made of $N_A \leq N$ consecutive sites, with $N_A/N=r$ and analyse the behaviour of $M_n^\Psi(r;\alpha)$. 

In Fig.~\ref{fig_Harmonic2} we compare results for two kinds of two-particle excited states: those of particles with identical charges and either distinct or equal momenta $p_1$ and $p_2$.
Our analytical predictions for $M_n(r;\alpha)$ are
\beqa
M_n^{1^{+} 1^{+}}(r;\alpha) &=& (r^n e^{2i\pi \alpha}+(1-r)^n)^2 \quad,\quad p_1\neq p_2\,, \nonumber \\
M_n^{2^{+}}(r;\alpha)&=&
r^{2n} e^{4 \pi i \alpha} + 2^n(1-r)^n r^n e^{2\pi i \alpha} + (1-r)^{2n} \quad,\quad p_1 = p_2\,.
\label{eq_HarmPred2}
\eeqa
In our numerics we have chosen $L = N =30$, so that the lattice spacing $L/N=1$. We also fix the mass scale to  $m = 0.1$, which corresponds to a typical correlation length of $\xi = m^{-1} = 10$ sites. Finally we choose either $p_1=p_2=\pi$ or $p_1=\pi$ and $p_2=\frac{2\pi}{5}$, both in units of the lattice spacing.
\medskip

Similarly, Fig.~\ref{fig_Harmonic3}, we consider the following three-particle excited states: a state of three equal momenta, that is $p_1 = p_2=p_3$, a state of two equal momenta among the three, that is $p_1=p_2\neq p_3$, and a state with three distinct momenta, that is $p_1,p_2,p_3$ distinct. In this case the analytical predictions are 
\beq 
M_n^{3^{+}}(r;\alpha) = r^{3n} e^{6i\pi \alpha}+3^nr^{2n}(1-r)^n e^{4i\pi \alpha}+ 3^nr^{n}(1-r)^{2n} e^{2i\pi \alpha}+(1-r)^{3n} \quad , \quad p_1=p_2=p_3\,, 
\eeq 
and 
\beqa
\begin{split}
M_n^{2^{+} 1^{+}}(r;\alpha)&=
M_n^{2^{+}}(r;\alpha) (r^n e^{2i\pi \alpha}+(1-r)^n) \quad, \quad p_1 = p_2\neq p_3\,,\\
M_n^{1^{+} 1^{+} 1^+}(r;\alpha)&=
(r^n e^{2i\pi \alpha}+(1-r)^n)^3 \quad,\quad p_1 \neq p_2 \neq p_3\,.
\end{split}
\label{eq_HarmPred3}
\eeqa
The set of momenta is $p_1=p_2=p_3=\pi$ for the first excited state, $p_1 = p_2 = \pi, p_3 = \pi/3$ for the second state, and $p_1 = \pi,p_2 = \pi/3, p_3 = \pi/5$ for the third one.

In all our figures we chose non-negative values of $\alpha$. Given the formulae above, taking $\alpha<0$ is equivalent to complex conjugation with $\alpha$ positive, so the figures for negative $\alpha$ are identical except for a change of sign in the imaginary part of all functions. We have also considered the value $\alpha=0$ (in green) which is the limit where there is no flux. 
As expected, in this case our formulae recover those for the excess Renyi Entropies 
in \cite{excited,excited2}, which are symmetric in $r$ and have vanishing imaginary part. Despite the fact that the correlation length is not particularly small with respect to the system's size $L$  ($\xi \simeq 0.33 L$), we took highly energetic states (momenta being fixed in the  large-volume limit) and we thus expect the validity of our predictions.

In both Fig.~\ref{fig_Harmonic2} and \ref{fig_Harmonic3},  we plot the numerical data (triangles) against analytical predictions (\eqref{eq_HarmPred2} and \eqref{eq_HarmPred3}) as functions of $r$ fixing $n=2$ for several values of $\alpha$ between $0$ and $\frac{1}{2}$, which correspond to flux $\pm 1$, respectively. At these two points, the ratio becomes purely real. The figures show excellent agreement between numerical data and analytical predictions.

\section{Magnon States}
\label{magnon}

The numerical results of the previous section provide convincing evidence that the results of \cite{partI} are correct. We now consider how our results might be applicable in a broader context. A natural starting point are  magnon states. Such states describe the eigenstates of a variety of spin chain Hamiltonians, with or without interactions. They admit a simple explicit form in the spin basis so that entanglement computations are easy to perform. We also know from \cite{excited2,ali6} that the total entanglement entropy of magnon states is described by our formulae with $\alpha=0$. As we see below, even in the presence of non-trivial scattering, the agreement extends to $\alpha\neq 0$. 

The main idea behind this construction is somewhat similar in spirit to the qubit picture \cite{excited2,partI}, namely, that the entanglement content of quasiparticles can be easily understood if one factors out the zero-point fluctuations. In other words, instead of considering the full quantum theory where the quasiparticles are constructed on top of a nontrivial ground state, which in general has its own entanglement content, we consider a simpler theory in which particles are constructed above a trivial ground state. It turns out that the entanglement of this simpler model keeps track of the exact entanglement of the quasiparticle and discards explicitly the entanglement of the true ground state.
Our magnon states belong to a Fock space generated by multiparticle configurations. In this case, the symmetry we construct will be ``internal", meaning that it acts just as a phase on each multiparticle state; this phase is directly related to the quantum numbers of each particle.

 \subsection{One-Magnon States}
We firstly focus on a single magnon state on the lattice, belonging to the one-particle sector of a quantum spin-$\frac{1}{2}$ chain of length $L$. A one-magnon state can be written as
\be
|\Psi_1\ket = \frac{1}{\sqrt{L}} \sum^L_{j=1}e^{ipj}|j\ket\,,
\ee
where $|j\ket$ is the state of a localised magnon of momentum $p$ in position $j$. If one imposes boundary conditions on the chain, the momentum $p$ is quantised as follows
\be
p \in \frac{2\pi}{L}\mathbb{Z}\,.
\ee
We introduce the action of the symmetry operator $e^{ 2\pi i \alpha Q}$, where $Q$ is associated with an internal symmetry. For our purposes we just need to specify its action on the vacuum state $|0\ket$ and on the one-particle sector. In addition we assume that the magnon is charged with respect to $Q$, and it has charge $+1$.
 
We are interested in the entanglement between spins $1,2,\cdots, \ell$ in region $A$ and the rest of the system. 
Associated to this region there is a restricted symmetry generator $e^{2\pi i\alpha Q_A}$, which acts as
\be
\begin{split}
e^{i2\pi\alpha Q_A}|0\ket = |0\ket\,, \qquad e^{2\pi i\alpha Q_A}|j\ket = e^{2\pi i\alpha \delta_{j\in A}}|j\ket\,,
\end{split}
\ee
where $\delta_{j\in A}$ gives 1 if $j$ is in $A$ and $0$ otherwise. The reduced density matrix of the region $A$ is
\be
\rho_A \equiv \text{Tr}_{\bar{A}}\l |\Psi_1\ket \bra \Psi_1 |\r = \frac{1}{L}\sum_{j,j' \in A}e^{ip(j-j')}|j\ket\bra j' | + (1-r)|0\ket\bra 0 |.
\ee
These two terms appearing in the formula above are interpreted as the contributions associated to the presence/absence of magnon in subsystem $A$, respectively. It is easy to show that
\be
e^{2\pi i \alpha Q_A}\rho_A^{n} = \l \frac{1}{L}\sum_{j,j' \in A}e^{ip(j-j')}|j\ket\bra j' | \r^n e^{2\pi i\alpha}+(1-r)^n |0\ket\bra 0 |,
\ee
 a relation which shows explicitly the presence of the symmetry for this state ($[\rho_A,e^{2\pi i\alpha Q_A}]=0$). After a straightforward calculation one gets
\be
\text{Tr}_A \l \frac{1}{L}\sum_{j,j' \in A}e^{ip(j-j')}|j\ket\bra j' | \r^n = r^n, \quad \text{Tr}_A\l (1-r)^n |0\ket\bra 0 |\r = (1-r)^n.
\ee
Putting the previous two pieces together, we arrive at the expected final result
\be
\text{Tr}\l \rho^n_A e^{2\pi i \alpha Q_A} \r = r^ne^{2 \pi i \alpha} + (1-r)^n\,,
\ee
with $r= \ell/L$, which provides the exact charged moments of a single magnon state.

\subsubsection{Two-Magnon States}

In the following we consider a state of two magnons with the same symmetry charge. This example is more interesting because it allows us to test whether the presence of non-trivial interaction changes our results. 
Given a pair of quasimomenta $p$ and $p'$, we parametrise this state in the following way
\be
|\Psi_2\ket = \frac{1}{\sqrt{L}}\sum_{j,j'}^L S_{j,j'}e^{ipj+ip'j'}| j j' \ket,
\ee
where $S$ is a scattering matrix and $| j j' \ket$ is the state with two localised magnons in sites $j$ and $j'$. The choice of the $S$-matrix is not really relevant for our purpose, but for the sake of concreteness we set
\be
S_{jj'} = \begin{cases} e^{i\varphi} \text{ for } j>j', \\ 1 \text{ for } j<j', \\ 0 \text{ for } j=j', \end{cases}
\ee
using the same conventions as in \cite{excited}. The action of the restricted symmetry operator $e^{2\pi i\alpha Q_A}$ on the two-particle sector of the Hilbert space is
\be
e^{2\pi i\alpha Q_A} | j j' \ket = e^{2\pi i \alpha (\delta_{j\in A}+\delta_{j'\in A})} | j j' \ket.
\ee
It is possible to decompose $\rho_A \equiv \text{Tr}_{\bar{A}}\l |\Psi_2\ket \bra \Psi_2 |\r$ as follows
\be
\rho_A = \frac{1}{L}\l \rho_A^{(1)} + \rho_A^{(2)} +\rho_A^{(3)}\r,
\ee
where $\rho_A^{(1)}$ is the two-particle contribution (both particles in $A$), $\rho_A^{(2)}$ is the vacuum contribution (no particles in $A$) and $\rho_A^{(3)}$ is the one-particle contribution (one particle in $A$ and one in $\bar{A}$). The introduction of the flux gives rise to the following relation
\be
\rho_A^n e^{2\pi i \alpha Q_A} = \frac{1}{L^n}\l (\rho_A^{(1)})^ne^{4\pi i\alpha} + \l \rho_A^{(2)}\r^n +\l\rho_A^{(3)}\r^n e^{2\pi i \alpha}\r.
\ee
No approximation was made up to this point, but the explicit expressions of $\rho_A^{(j)}$, given in \cite{excited}, are cumbersome and not particularly enlightening for our purpose. However, one can show that in the limit $L\rightarrow \infty$ and $\ell/L,p\neq p'$ kept fixed, $\text{Tr}_A\l (\rho_A^{(j)})^n\r$ simplifies drastically:
\be
\text{Tr}_A\l (\rho_A^{(1)})^n\r \simeq L^n r^{2n}, \quad \text{Tr}_A\l (\rho_A^{(2)})^n\r \simeq L^n (1-r)^{2n}, \quad \text{Tr}_A\l (\rho_A^{(3)})^n\r \simeq 2L^n r^n(1-r)^{n}.
\ee
Putting all the pieces together one finally gets
\be
\text{Tr}_A\l \rho_A^n e^{2\pi i \alpha Q_A} \r \simeq r^{2n}e^{4\pi i\alpha} +2r^n(1-r)^ne^{2\pi i\alpha}+(1-r)^{2n} = (r^ne^{2\pi i\alpha}+(1-r)^n)^2.
\ee
This computation shows that in this particular scaling limit the interaction between particles has no effect on the final result, and the total charged moment is just a product of two single-particle charged moments.
A different result is obtained if $p=p'$ and fixed. In that case, the magnons are indistinguishable and one can prove that
\be
\text{Tr}_A\l (\rho_A^{(1)})^n\r \simeq L^n r^{2n}, \quad \text{Tr}_A\l (\rho_A^{(2)})^n\r \simeq L^n (1-r)^{2n}, \quad \text{Tr}_A\l (\rho_A^{(3)})^n\r \simeq 2^nL^n r^n(1-r)^{n},
\ee
so that
\be
\text{Tr}_A\l \rho_A^n e^{2\pi i\alpha Q_A} \r \simeq r^{2n}e^{4\pi i\alpha} +2^n r^n(1-r)^n e^{2\pi i\alpha}+(1-r)^{2n},
\ee
which no longer factorises into one-magnon contributions. Both results are special cases of (\ref{una}) and (\ref{for1}).

The results of this section generalise previous work for the excess entanglement entropy of excited states \cite{excited2} and are also related to the results of \cite{ali6} where the entanglement of magnon states was considered more generally. In particular, it was shown that for states consisting of several magnons, entanglement will factorise into the contributions of groups of magnons which are well-separated from each other in momentum space (that is their momentum difference is of order $1$ rather than $1/L$, where $L$ is the length of the system). Such results also apply to the present case up to the introduction of the appropriate phases.

\section{Further Generalisations: Higher Dimensions}
\label{higher}

So far we have derived the behaviour of the charged moments of the SREE of quasiparticle excited states making use of two different formalisms: the form factor expansion in 1+1 integrable QFTs \cite{partI} and the analysis of qubit/magnon states on the lattice. Unfortunately these techniques are specially suited for 1+1D theories. In this section we want to consider instead a generic QFT in higher dimensions. To this aim we introduce a slightly different approach. Indeed, on the one hand, the description of the R\'enyi entropy as a correlation functions of branch point twist fields inserted at some points of the space is special of 1+1 dimensional QFT \cite{Calabrese:2004eu,entropy}. On the other hand, the description of excited states as magnon/qubit states, even if generalisable to higher dimensions, has the disadvantage that it does not take into account the zero-point fluctuations.

Despite these technical limitations, we expect that in the particular scaling limit we are considering the universal entanglement content of the R\'enyi entropy, together with its symmetry resolved version, should not depend on dimensionality, presence or not of interactions or even integrability.  To support these claims, Section 3 on magnonic states provided evidence that the same results tested numerically for free theories in Section 2 can be recovered for interacting systems. As for higher dimensions, at least one precedent for this generalisation already exists. In a previous work \cite{excited4} the excitations of the free massive boson in $D:=d+1$ dimensions have been extensively analysed and their R\'enyi entropy was computed in terms of graph partition functions, and found once more to fit the same formulae, with $r$ replaced by the ratio of generalised volumes. 

In this section we slightly generalise the formalism of \cite{excited4}, to take into account possible interactions, and provide, as a proof of concept, a simple calculation of symmetry resolved entanglement of a single-particle excited state. The key ingredients we need are the description of the excited states as local operators acting on a vacuum state and a twist field operator, which generalizes the composite branch point twist field to higher dimensional settings (an introduction to composite twist fields can be found in \cite{partI}). The only strong assumption we make in our derivation is the presence of a finite mass gap $m$, whose inverse $m^{-1}$ is much smaller than the typical lengths of the system.\footnote{This is probably not necessary, since the emergence of the universal entanglement content is also expected for some high-energy states in massless theories (see \cite{ali4,ali6}, for the analysis of the gapless XY chain). However, we keep this assumption here mostly to avoid technical complications, leaving the analysis of CFTs to future investigations.}

We anticipate here that our formulae (\ref{una})-(\ref{general}) are unchanged in higher dimensional theories, up to the identification
\be
r = \frac{V_A}{V},
\label{higherr}
\ee
which is the ratio of (generalised) volumes between subsystem $A$ and the total system.

\medskip
 It may seem surprising that results should only depend on $r$, as defined in (\ref{higherr}), and not on other features of the entanglement region, such as the connectivity and smoothness of its boundary. Indeed, the charged moments and symmetry resolved entropies of both the ground state and of excited states will depend on such properties, as would finite volume corrections to our results. However, our computations deliver results for the ratio of charged moments between the excited and ground states in the infinite-volume limit, and it is this ratio in this limit which is universal and independent of boundary features, not the charged moments themselves. This independence of boundary features has been analytically shown for one simple example, namely the case of one-dimensional disconnected regions, where the same formulae as for one connected region were found to apply, with $r$ the sum of the lengths of all disconnected parts \cite{excited2}.  

\subsection{Excited States and Operator Algebra}
 Let us consider the vacuum state $|0\ket$ of a Hilbert space $\mathcal{H}$, together with an algebra $\mathcal{A}$ of observables\footnote{In the case of a single real boson, $\mathcal{A}$ is just the algebra of operators generated by the field $\Phi(x)$ and its conjugated momentum $\Pi(x)$.} acting on $\mathcal{H}$ which has $|0\ket$ as a cylic vector (see \cite{Witten-18} for a modern review of this algebraic viewpoint in QFT). This allows us to represent any state $|\Psi\ket$ of the Hilbert space as
\be
|\Psi\ket = \mathcal{O}|0\ket \quad \mathrm{with} \quad \mathcal{O} \in \mathcal{A}\,.
\ee
We would like to assume further that the vacuum state is translation invariant, namely that it is invariant under a certain faithful representation of the translation group in $d$ dimensions. Strictly speaking, since we consider a finite-size system, we have to slightly modify this requirement. Specifically, we put our system on a $d$-dimensional torus $\mathcal{M}$ of volume $V$ and we require that $|0\ket$ is invariant under the isometries of the torus. Other boundary conditions can be considered too, but they do not change the picture in the scaling limit we are interested in. We also require \text{locality} of the observables, asking that $\mathcal{A}$ is generated by a set of fields $\{\mathcal{O}(\mathbf{x})\}$, which can be applied at any point of $\mathcal{M}$.

For any field $\mathcal{O}(\mathbf{x})$ one can construct its Fourier transform $\mathcal{O}(\mathbf{p})$ as
\be
\mathcal{O}(\mathbf{p}) = \int_{\mathcal{M}}d^dx e^{-i\mathbf{px}}\mathcal{O}(\mathbf{x})\,,
\ee
and these transformed fields are building blocks for the following set of translation invariant states
\be
\mathcal{O}_1(\mathbf{p}_1)\dots \mathcal{O}_k(\mathbf{p}_k)|0\ket\,.
\ee
The state above corresponds physically to $k$ particles distributed on $\mathcal{M}$ with momenta $\{\mathbf{p}_j\}_{j=1,\dots,k}$, and the choice of the fields $\{\mathcal{O}_j\}$ may depend on the particle species and quantum numbers. At first sight, this construction is similar to the usual way of generating particle states in free theories acting with creation operators on the vacuum on a Fock space. However, the real advantage of our formulation is that it is directly related to local observables, a property which is fundamental to correctly define entanglement measures.

Let us take a set of orthogonal fields $\{\mathcal{O}_j\}$, so that the correlation function $\la \mathcal{O}^\dagger_i(\mathbf{x})\mathcal{O}_j(\mathbf{x}') \ra$ vanishes for $i\neq j$. In other words the fusion
\be
\mathcal{O}^\dagger_i \times \mathcal{O}_j \rightarrow 1
\ee
is present only if $i=j$ and their operator product expansion (OPE) can be expressed formally as
\be
\mathcal{O}^\dagger_i(\mathbf{x})\mathcal{O}_j(\mathbf{x}') \simeq  \delta_{ij} \bra 0 | \mathcal{O}^\dagger_i(\mathbf{x}-\mathbf{x}')\mathcal{O}_i(0) | 0\ket + \dots,
\label{OPE_op}
\ee
where we neglected explicitly the contributions coming from less relevant operators (with respect to the identity). The exact evaluation of the correlation function above
can be hard, but the assumption of a finite gap $m$ ensures that it vanishes exponentially for $|\mathbf{x}-\mathbf{x}'| \gg m^{-1}$. This is the only property we really need in our subsequent discussion.

Consider a {smeared} version of the modes $\mathcal{O}(\mathbf{p})$, with support in a subsystem only, that is a region of space. To each spacial region $A \subseteq \mathcal{M}$
and field $\mathcal{O}(\mathbf{x})$, we associate
\be
\mathcal{O}_A(\mathbf{p}) = \int_{A}d^dx e^{-i\mathbf{px}}\mathcal{O}(\mathbf{x}).
\ee
Given any two regions $A$ and $A'$, we compute\footnote{One should note that hermitian conjugation and Fourier transform do not commute. Indeed, we have that $\mathcal{O}^\dagger_A(-\mathbf{p}) = \l\mathcal{O}_A(\mathbf{p})\r^\dagger$.} $\mathcal{O}^\dagger_A(-\mathbf{p})\mathcal{O}_{A'}(\mathbf{p}')$, making use of some approximations. First, we consider only the most relevant term in the fields OPE, namely
\be
\mathcal{O}^\dagger_A(-\mathbf{p})\mathcal{O}_{A'}(\mathbf{p}') \simeq \int_A d^dx \int_{A'} d^dx' e^{i\mathbf{px}-i\mathbf{p}'\mathbf{x}'} \bra 0 | \mathcal{O}^\dagger(\mathbf{x})\mathcal{O}(\mathbf{x}')|0\ket\,.
\ee
Second, since we are working in the limit of small correlation length (compared to the geometry), the leading contribution comes from the insertion of the fields at small distances, which is present if $\mathbf{x},\mathbf{x}' \in A \cap A'$; this observation motivates the change of variable $\mathbf{x}'' = \mathbf{x}'-\mathbf{x}$, and the subsequent approximation
\be
\mathcal{O}^\dagger_A(-\mathbf{p})\mathcal{O}_{A'}(\mathbf{p}') \simeq \int_{A\cap A'} d^dx e^{i(\mathbf{p}-\mathbf{p}')\mathbf{x}} \cdot \int_\mathcal{M}dx'' e^{-i\mathbf{p}'\mathbf{x}''}\bra 0 | \mathcal{O}^\dagger(0)\mathcal{O}(\mathbf{x}'')|0\ket
\label{eq:OPE_Ftransform}
\ee
The second integral may be difficult to compute and in principle it could require a UV regularisation for $|\mathbf{x}''| < \epsilon \ll m^{-1}$. However it does not depend on the regions $A,A'$ and in our computation appears only as a multiplicative constant. In conclusion, we end up with
\be
\mathcal{O}^\dagger_A(-\mathbf{p})\mathcal{O}_{A'}(\mathbf{p}') \propto V_{A \cap A'} \delta_{\mathbf{p},\mathbf{p}'},
\label{OPE_four}
\ee
where $V_{A \cap A'}$ is the volume of $A \cap A'$, which is the main result of this subsection. Since the volume in (\ref{OPE_four}) emerges from the integrals (\ref{eq:OPE_Ftransform}), which involve a Fourier transform, we require that subsystem $A\cap A'$ consists of a finite number of disconnected regions, whose boundaries are piecewise smooth. 
\medskip

It is natural to ask how our discussion above would be modified for a vanishing gap $m=0$. The main change is in the scaling of correlations functions: exponential localization of the correlation function in a region of typical length $m^{-1}$ does not hold any longer, due to the long algebraic tails of the correlation functions. We conjecture that, as long as the momenta are fixed in the infinite-volume limit, the main conclusion \eqref{OPE_four} is unchanged. A qualitative argument is that in this case the inverse momentum, say the De Broglie length, plays the role of typical length scale. In order to make this consideration more precise, let us analyse Eq. \eqref{eq:OPE_Ftransform} for a 1+1D CFT, where $\mathcal{O}$ is a field of conformal dimension $\Delta_{\mathcal{O}}$. We focus on the following integral
\be
\int_\mathcal{M}dx'' e^{-i\mathbf{p}'\mathbf{x}''}\bra 0 | \mathcal{O}^\dagger(0)\mathcal{O}(\mathbf{x}'')|0\ket,
\ee
which we regulate both in the UV, with a cutoff $\epsilon$, and in the IR, with a cutoff $L$, as follows
\be
\int^{L}_{\epsilon} dx e^{-ipx} \frac{1}{x^{4\Delta_\mathcal{O}}} + (\text{c.c.}).
\label{eq:Integral_1D}
\ee
This integral can be explicitly computed. However, the important feature is that for $\Delta_\mathcal{O}>0$, $p>0$ and $\epsilon>0$ all fixed, the integral converges to a finite value when $L\rightarrow +\infty$. This is no longer the case if  $p\sim 1/L$ in the infinite-volume limit. In practice, this means that for small momentum and scaling dimension $0<\Delta_\mathcal{O}\leq 1$ the considerations we made so far regarding the scaling at large sizes cannot be applied. As a matter of fact, for free CFTs the scaling dimensions of the fundamental fields are smaller than $1$:  the fermionic field $\Psi$ has dimension $1/2$ while the derivative of a compact boson $\partial_x \Phi$ has dimension $1$.

\medskip 
While these considerations are not mathematically rigorous in establishing convergence of the OPE expansion in the large volume limit, they are sufficient to explain why low-energy states of gapless theories, or multiparticle states with small momenta difference, are not well captured by our predictions. Indeed, for such states the excess entanglement was computed in \cite{german1, german2} and is clearly different from the formulae in \cite{excited,excited2}.

\subsection{Replica Construction for Symmetry Resolved Entanglement}

Consider now a replica version of the theory, consisting of $n$ copies. For any state $|\Psi \ket$ we consider its replicated version $|\Psi \ket^n$. Our goal is to define a composite twist operator, which generalises the composite branch point twist field as defined in \cite{GS}, generalising \cite{entropy}, to higher dimensional theories. It is well known from a large body of literature e.g. \cite{GS,Horvath_2021,FFSRE,horvath2021branch,pottsSRE} that the charged moments of the symmetry resolved R\'enyi entropies are given by correlation functions of composite twist fields (CTF). A short review of their main properties can be found in \cite{partI}.

In this section, we want to extend the notion of CTFs to higher dimensions. To this aim we will introduce a composite twist operator whose expectation value over $|\Psi \ket^{n}$ gives exactly the charged moments $Z_n(\alpha)$. This type of operator was already considered in the literature, especially in the absence of the flux insertion (see for example \cite{excited4,hms-14,Sveskoa-21,Long-21}), but here we are mostly interested in its relationship with the algebra of local operators.

The first point we have to clarify regards the symmetry, and its action on the space of fields. Starting from $e^{2\pi i \alpha Q}$, the global generator of $U(1)$ symmetry in the non-replicated theory, we say that $\mathcal{O}(\mathbf{x})$ has charge $\kappa_\mathcal{O}$ if
\be
e^{2\pi i\alpha Q} \mathcal{O}(\mathbf{x}) e^{-2\pi i \alpha Q} = e^{2\pi i \alpha \kappa_\mathcal{O} } \mathcal{O}(\mathbf{x}).
\ee
Since one can decompose the space of fields as irreducible representations of $U(1)$, we restrict our analysis to charged fields. Going back to the replicated theory, we consider the algebra of replicated observables $\mathcal{A}^{n}$ as the algebra generated by the tensor product of $n$ observables in $\mathcal{A}$. Thus, to any field $\mathcal{O}(\mathbf{x}) \in \mathcal{A}$, we associate $\mathcal{O}^j(\mathbf{x}) \in \mathcal{A}^{n}$ defined as
\be
\mathcal{O}^j(\mathbf{x}) = 1 \otimes \cdots \otimes 1 \otimes  \mathcal{O}(\mathbf{x}) \otimes 1 \cdots 1,
\ee
where $\mathcal{O}(\mathbf{x})$ lives only in the $j$th replica. 
Consider a spacial region $A$, and its complement $\bar{A}$. We define a composite twist operator $T^\alpha_A$ which implements the structure of the $n$-sheeted, cyclically connected, Riemann surface where the replica theory is defined. That is $T_A^\alpha$ ``implements" the gluing of the replicas along $A$ with an additional flux insertion due to the action of the $U(1)$ symmetry. As for a standard CTF, it does so via its commutation relations with any charged field $\mathcal{O}^j(\mathbf{x})$, which we require to be
\be
T^\alpha_A \mathcal{O}^j(\mathbf{x}) = \begin{cases} e^{2\pi i \kappa_\mathcal{O} \alpha \delta_{j,n}} \mathcal{O}^{j+1}(\mathbf{x}) T^\alpha_A \quad \mathbf{x} \in A, \\
\qquad \qquad \quad \,\mathcal{O}^{j}(\mathbf{x}) T^\alpha_A \quad \mathbf{x} \in \bar{A}.
\end{cases}
\label{comm_rel}
\ee
Using this choice, the flux is inserted only between the $n$th and the first replica. We would like to emphasize that a similar definition has already appeared in the context of 1+1D integrable QFTs (see \cite{Horvath_2021,FFSRE,horvath2021branch,pottsSRE
}). In particular for $A = [0,\ell]$, one can identify
\be
T^\alpha_A = \mathcal{T}_n^\alpha(0)\tilde{\mathcal{T}}_n^\alpha(\ell),
\ee
and the commutation relations for $T^\alpha_A$ can be expressed as commutation relations for the CTF $\mathcal{T}_n^\alpha$. The only novelty here is that for higher-dimensional theories it is not clear what the replacement for CTFs is, thus how to represent  $\mathcal{T}_n^\alpha$. Regarding the definition of the twist operator $T^\alpha_A$, it is worth mentioning also that in QFT it is known\cite{dvvv-89,dei-18} that the twist operators are not local observables of the algebra $\mathcal{A}^{n}$. They are rather observables of the orbifolded algebra usually written as $\mathcal{A}^{n}/\mathbb{Z}_n$, which extends $\mathcal{A}^{n}$ and is obtained by taking the coset over the cyclic symmetry of replicas.

We are now ready to relate the twist operator to the symmetry resolved entanglement. The charged moments of $|\Psi\ket$ are given by
\be
Z^\Psi_n(\alpha) = \frac{^n\bra \Psi| T^\alpha_A|\Psi \ket^{ n}}{^{ n}\bra \Psi |\Psi \ket^{n}}.
\ee
The above definition, together with the commutation relations \eqref{comm_rel} and the OPE of Eq. \eqref{OPE_four}, should be enough to prove the explicit analytical expression of the ratio of charged moments between $|\Psi\ket$ and the ground state defined in (\ref{eq_Mratio}). We now show how these ideas come together for a simple example.

\subsection{Single-Particle State}

In this section we analyse an excited state $|\Psi\ket$ made of a single quasiparticle with momentum $\mathbf{p}$ generated by a charged field $\mathcal{O}$. Its explicit expression is given by
\be
|\Psi\ket = \mathcal{O}(\mathbf{p})|0\ket,
\ee
and the replicated version is just
\be
| \Psi\ket^{n} = \mathcal{O}^1(\mathbf{p})\dots \mathcal{O}^n(\mathbf{p})|0\ket^n.
\ee
For the sake of convenience, we split
\be
\mathcal{O}^j(\mathbf{p}) = \mathcal{O}^j_A(\mathbf{p}) + \mathcal{O}^j_{\bar{A}}(\mathbf{p}),
\ee
so that its commutation relations with $T^\alpha_A$ become more transparent. Indeed, using just \eqref{comm_rel} one can express
\be
\begin{split}
T^\alpha_A| \Psi\ket^{ n} = T^\alpha_A(\mathcal{O}^1_A(\mathbf{p})+ \mathcal{O}^1_{\bar{A}}(\mathbf{p}))\dots (\mathcal{O}^n_A(\mathbf{p}) + \mathcal{O}^n_{\bar{A}}(\mathbf{p}) ) |0\ket^{ n} =\\
(\mathcal{O}^2_A(\mathbf{p})+ \mathcal{O}^1_{\bar{A}}(\mathbf{p}))\dots (\mathcal{O}^1_A(\mathbf{p})e^{2\pi i \alpha \kappa_\mathcal{O}} + \mathcal{O}^n_{\bar{A}}(\mathbf{p}) ) T^\alpha_A |0\ket^{ n}.
\end{split}
\ee
 Up to now everything looks exact, while the approximate evaluation of $^{n}\bra \Psi |T^\alpha_A| \Psi\ket^{ n}$ makes use of the OPE contraction in \eqref{OPE_four}. Note also how the phase $e^{2\pi i \alpha}$ is only present for $n$th factor above, similar to the Gaussian measure found by employing wave-functionals in Section 2. Among all the terms which are generated, all but two are vanishing and they give
\be
\begin{split}
^{ n}\bra \Psi |T^\alpha_A| \Psi\ket^{ n} \simeq e^{2\pi i \alpha \kappa_\mathcal{O}} \ ^{ n}\bra 0 |(\mathcal{O}^\dagger)^n_A(-\mathbf{p})\ldots(\mathcal{O}^\dagger)^1_A(-\mathbf{p}) \mathcal{O}^2_A(\mathbf{p})\dots \mathcal{O}^n_A(\mathbf{p}) \mathcal{O}^1_A(\mathbf{p})T^\alpha_A|0\ket^{ n}   \\
+ ^{ n}\bra 0 |(\mathcal{O}^\dagger)^n_{\bar{A}}(-\mathbf{p})\ldots(\mathcal{O}^\dagger)^1_{\bar{A}}(-\mathbf{p}) \mathcal{O}^1_{\bar{A}}(\mathbf{p})\mathcal{O}^2_{\bar{A}}(\mathbf{p})\dots \mathcal{O}^n_{\bar{A}}(\mathbf{p})T^\alpha_A|0\ket^{ n}\\
\propto \l e^{2\pi i\alpha \kappa_\mathcal{O}} V^n_A + (V-V_A)^n\r \frac{^{ n}\bra 0 | T^\alpha_A | 0 \ket^{ n}}{^{ n}\bra 0 | 0 \ket^{ n}}.
\end{split}
\ee
Similarly, we can evaluate the norm $^{ n}\bra \Psi |\Psi\ket^{ n}$ which does not require the splitting of $\mathcal{O}^j(\mathbf{p})$
\be
^{ n}\bra \Psi| \Psi\ket^{n} = \\
^n\bra 0 |(\mathcal{O}^\dagger)^n(-\mathbf{p})(\mathcal{O}^\dagger)^1(-\mathbf{p}) \mathcal{O}^1(\mathbf{p})\dots \mathcal{O}^n(\mathbf{p})|0\ket^n\propto V^n.
\ee
In the evaluation of the ratio
\be
\frac{^{ n}\bra \Psi |T^\alpha_A| \Psi\ket^{n}}{^{ n}\bra \Psi| \Psi\ket^{ n}}
\ee
the proportionality constant (which is non-universal and could be absorbed in a redefinition of the field) cancels out, and one can  write
\be
\frac{^{ n}\bra \Psi |T^\alpha_A| \Psi\ket^{ n} }{^{ n}\bra \Psi| \Psi\ket^{ n}} \simeq  \l e^{2\pi i\alpha \kappa_\mathcal{O}} r^n + (1-r)^n\r \frac{^n\bra 0 | T^\alpha_A | 0 \ket^n}{^n\bra 0 | 0 \ket^n}
\ee
with $r = \frac{V_A}{V}$. In the expression above, the first piece is universal while the second  is not, it is just the $n$th charged moment of the ground-state. Taking the ratio with the ground state contribution, we finally arrive to the desired result
\be
M_n^\Psi(r,\alpha) = \frac{^{ n}\bra \Psi |T^\alpha_A| \Psi\ket^{ n} }{^{ n}\bra \Psi| \Psi\ket^{ n}} \frac{^{ n}\bra 0| 0\ket^{ n}} {^{ n}\bra 0 |T^\alpha_A| 0\ket^{ n} } \simeq  e^{2\pi i \alpha \kappa_\mathcal{O}} r^n + (1-r)^n.
\ee
Results for multiparticle states can be obtained in a similar fashion. 
\medskip

To conclude, the striking simplicity of these results, and the possible generalisations to multiparticle states, rely especially on the truncation of the OPE in \eqref{OPE_op}, which is expected to become exact in the limit $m^d V\gg 1$. 
We further expect that for finite $m^dV$ further contributions in the OPE can be recast as a (possibly non-integer) power series in $({m^d V})^{-1}$, which generalises the explicit $(mL)^{-1}$ power expansion that is obtained for 1+1D free theories using form factor techniques \cite{partI}. In massless theories, we instead expect corrections as a power series in $(|\mathbf{p}|^dV)^{-1}$.

The explicit evaluation of these corrections, which are expected to be non-universal, that is, momentum and QFT-dependent, and any possible issues regarding the convergence of these power series are all beyond the scope of this work.

\section{Conclusions}
\label{conclusion}

In this paper we have extended the results of \cite{partI} by first, providing numerical evidence for their validity and second, showing that they apply much more broadly than the form factor computation of \cite{partI} would suggest, to interacting and higher dimensional theories. 
\medskip

Regarding the numerics, we have considered a 1D Fermi gas and a complex harmonic chain. Although both these models are discrete and amenable to numerical computations, their microscopic features are extremely different. Whereas the 1D Fermi gas is a fermionic theory whose continuous limit is a massless complex free fermion, the complex harmonic chain's continuous limit is a complex massive free boson. It is therefore quite remarkable that the same set of formulae for the ratios of charged moments should apply for both theories. This is nonetheless the case. Whereas for the 1D Fermi gas our formulae hold for highly excited states containing excitations of large momenta, thus small De Bloglie wavelengths, for the complex harmonic charge, the formulae hold as long as the correlation length is small compare to subsystem size and the excited states are usually of low energy. The common feature of both types of states is that they are characterised by localised excitations. 

A simple class of states in one space dimension are magnon states. These are typical excited states of quantum spin chains, with and without interaction. They have a simple representation in terms of spin degrees of freedom and their entanglement entropies can be computed analytically. We have shown that, irrespective of interaction, the formulae found in \cite{partI} still apply, in line with observations made in \cite{excited2,ali6}. 

Finally, we have proposed a method to compute the charged moments of zero-density excited states in higher-dimensions with or without interactions. The generalisation relies on generalising the action of branch point twist fields to higher dimensions (without generalising the fields themselves) and on standard assumptions about the asymptotics of correlations functions of local fields. 

\medskip

There are several extensions of this work that are still outstanding: the study of finite-volume corrections to these results in a QFT setting, the extension of our formulae to other symmetry resolved measures of entanglement, such as the logarithmic negativity \cite{expSRE2,parezlatest,excited3}, and the study of theories where quasiparticle excitations are not localised (such as free fermion chains where the local degrees of freedom are the spins, which are non-local with respect to fermions). We hope to return to these problems in future work. 

\medskip

 \noindent {\bf Acknowledgements:} We would like to thank Benjamin Doyon for his invaluable help in developing the harmonic chain's numerical simulation. Luca Capizzi thanks ERC for support under Consolidator grant number 771536 (NEMO). Cecilia De Fazio thanks EPSRC for financial support under Grant EP/V031201/1. Michele Mazzoni is grateful for funding under the EPSRC Mathematical Sciences Doctoral Training Partnership EP/W524104/1. Luc\'ia Santamar\'ia-Sanz is grateful to the Spanish Government for funding under the FPU-fellowships program FPU18/00957, the FPU Mobility subprogram EST19/00616, and MCIN grant PID2020-113406GB-I0. Olalla A. Castro-Alvaredo thanks EPSRC for financial support under Small Grant EP/W007045/1.

\appendix

\section{Trace Calculations via Wave Functional Method}
\label{AppA}
In this Appendix we present explicit calculations of the ratio of charged moments for excited states in a complex free boson theory using the wave functional method introduced in Section \ref{waveF}. The discretisation of these results leads to the formulae for the complex harmonic chain presented in Subsection \ref{23}.

\subsection{Zero flux}
We now wish to compute the ratio of charged moments
\be
\frac{\Tr_A\left(\rho^n_{A}\,e^{2\pi i \alpha Q_A}\right)}{\Tr_A\left(\rho^n_{0,A}\,e^{2\pi i \alpha Q_A}\right)}\,,
\label{ratiof}
\ee
where $\rho_A$ and $\rho_{0,A}$  are the reduced density matrices of the excited and ground states, respectively.  

We define restricted wave functionals which take as arguments (complex) fields that either have support on region $A \equiv [0, \ell)$ or on  $\bar{A} \equiv [\ell , L]$. Leaving the dependence on the conjugate fields implicit, we write:
\begin{align}
\Phi(x){\boldsymbol{\Psi}} [\phi_A , \phi'_{\bar{A}}] &= \left(\delta_{x \,\in\, A}\phi(x) + \delta_{x\, \in \,{\bar{A}}}\phi'(x)\right){\boldsymbol{\Psi}} [\phi_A , \phi'_{\bar{A}}]\,, \\ i\Pi(x){\boldsymbol{\Psi}} [\phi_A , \phi'_{\bar{A}}] &= \left(\delta_{x\,\in \, A}\frac{\delta}{\delta \phi(x)}+\delta_{x\,\in \,{\bar{A}}}\frac{\delta}{\delta \phi'(x)}\right){\boldsymbol{\Psi}} [\phi_A , \phi'_{\bar{A}}]\,,
\end{align}
and similarly for the action of $\Phi^\dagger(x)$, $i\Pi^\dagger(x)$. The action of the charge operator $Q_A$ on the wave functional is simply:
\be 
Q_A {\boldsymbol{\Psi}} [\phi_A , \phi'_{\bar{A}}] = \int_A \mathrm{d}x\,\left(\phi^\dagger(x)\frac{\delta}{\delta \phi^\dagger(x)} - \phi(x)\frac{\delta}{\delta \phi(x)}\right){\boldsymbol{\Psi}} [\phi_A , \phi'_{\bar{A}}]\,,
\ee
while from \eqref{Q_action_ket} it follows that
\be
e^{2\pi i \alpha Q_A}|\phi_A , \phi'_{\bar{A}}\ket = |e^{2\pi i \alpha}\phi_A , \phi'_{\bar{A}}\ket\,.
\ee
The reduced density matrix then admits the functional representation
\be
\rho_A = \Tr_{\bar{A}} \rho = \int \mathcal{D}\phi_{\bar{A}}  \bra \phi_{\bar{A}} | \rho | \phi_{\bar{A}} \ket 
\Rightarrow \bra \phi'_A | \rho_A |\phi''_A \ket = \int \mathcal{D} \phi_{\bar{A}} {\boldsymbol{\Psi}} [\phi'_A , \phi_{\bar{A}}] {\boldsymbol{\Psi}} [\phi''_A, \phi_{\bar{A}}]^*\,,
\ee
where $\mathcal{D}\phi$ is shorthand for $\mathcal{D}\phi\mathcal{D}\phi^\dagger$\,.

We start with the simple case where there is no symmetry resolution, that is, we compute (\ref{ratiof}) for $\alpha=0$, following \cite{excited2}. For $n$ integer, we can insert the resolution of the identity $n$ times, with the identification $n+1 \equiv 1$, and we get:
\begin{align}
\label{denominator_no_flux}
\Tr_A(\rho^n_{0,A}) &= \int \mathcal{D}\phi_{1A}\dots \mathcal{D}\phi_{nA} \bra \phi_{1A} |\rho_{\text{vac},A} | \phi_{2A} \ket \dots \bra \phi_{nA} |\rho_{\text{vac},A} |\phi_{1A} \ket \nonumber \\
&= \int \mathcal{D}\phi_1\dots \mathcal{D}\phi_n \prod_{i=1}^n {\boldsymbol{\Psi}}_\text{vac}[\phi_{i,A}, \phi_{i,\bar{A}}]\prod_{i=1}^n {\boldsymbol{\Psi}}_\text{vac}[\phi_{i+1,A}, \phi_{i,\bar{A}}]^*\,.
\end{align}
The product of the diagonal terms, i.e. those in which the fields act on the same copy for both subsystems is
\be
\prod_{i=1}^n {\boldsymbol{\Psi}}_\text{vac}[\phi_{i,A}, \phi_{i,{\bar{A}}}] = \exp \left[-\sum_{i=1}^n \int_{A \bigcup {\bar{A}}} \mathrm{d}x\mathrm{d}y \,\phi_i^\dagger(x)K(x-y)\phi_i(y)\right]\,,
\ee
with $K(x)$ defined in \eqref{vacuum_functional}.
For the non-diagonal terms we notice that since $K(x)^* = K(-x)$ we have ${\boldsymbol{\Psi}}_\text{vac}[\phi_{i+1,A}, \phi_{i,{\bar{A}}}]^* = {\boldsymbol{\Psi}}_\text{vac}[\phi_{i+1,A}, \phi_{i,{\bar{A}}}]$ and:

\begin{align}
&\prod_{i=1}^n {\boldsymbol{\Psi}}_\text{vac}[\phi_{i+1,A}, \phi_{i,{\bar{A}}}] = \exp\left\{ -\sum_{i=1}^n \left[\int_{\substack{x \, \in \, A \\ y \, \in \, A}}\mathrm{d}x\mathrm{d}y\,\phi_{i+1}^\dagger(x)K(x-y)\phi_{i+1}(y) \right. \right. \\ &\left. \left.+ \int_{\substack{x \, \in \, {\bar{A}} \\ y \, \in \, {\bar{A}}}}\mathrm{d}x\mathrm{d}y\,\phi_i^\dagger(x)K(x-y)\phi_i(y) + \left(\int_{\substack{x \, \in \, A \\ y \, \in \, {\bar{A}}}}\mathrm{d}x\mathrm{d}y\,\phi_{i+1}^\dagger(x)K(x-y)\phi_i(y) + \text{c.c.}\right) \right] \right \} \,.
\end{align}
Putting all the terms together, we end up with the Gaussian measure:
\be
\Tr_A(\rho^n_{0,A}) = \int \mathcal{D}\phi_1 \dots \mathcal{D}\phi_n \exp \left[-\mathcal{G}\right] \,, 
\ee
where
\beqa
\mathcal{G} &=& \sum_{i=1}^n  2\left(\int_{\substack{x \, \in \, A \\ y \, \in \, A}} + \int_{\substack{x \, \in \, {\bar{A}} \\ y \, \in \, {\bar{A}}}}\right)\phi_i^\dagger(x)K(x-y)\phi_i(y) \nonumber \\ &&
+  \sum_{i=1}^n\left(\int_{\substack{x \, \in \, A \\ y \, \in \, {\bar{A}}}} (\phi_i(x) + \phi_{i+1}(x))^\dagger K(x-y)\phi_i(y) + \text{c.c.}\right)\,.
\eeqa 
The numerator of (\ref{ratiof}) instead is
\be 
\Tr_A\left(\rho^n_{A}\right) = \int \mathcal{D}\phi_1\dots \mathcal{D}\phi_n \prod_{i=1}^n {\boldsymbol{\Psi}}_{\{p_{j^+}p_{j^-}\}}[\phi_{i,A}, \phi_{i,{\bar{A}}}]\prod_{i=1}^n {\boldsymbol{\Psi}}_{\{p_{j^+}p_{j^-}\}}[\phi_{i+1,A}, \phi_{i,{\bar{A}}}]^*
\ee 
for a state of $k^\pm$ particles of charges $\pm 1$ and momenta $p_{j^\pm}$ (assuming $p_{j^\pm} \ne p_{i^\pm}$ for $j^\pm \ne i^\pm$), and using \eqref{multi-particle state}:
\begin{align}
\prod_{i=1}^n {\boldsymbol{\Psi}}_{\{p_{j^+}p_{j^-}\}}[\phi_{i,A}, \phi_{i,{\bar{A}}}] &= \left(\prod_{j^+}\frac{2E_{p_{j^+}}}{L}\prod_{j^-}\frac{2E_{p_{j^-}}}{L}\right)^\frac{n}{2}\prod_{i=1}^n\prod_{j^+}\int_{A \bigcup {\bar{A}}}\mathrm{d}x e^{ip_{j^+}x}\phi_i^\dagger(x)\nonumber \\ &\times\prod_{j^-}\int_{A \bigcup {\bar{A}}}\mathrm{d}x e^{ip_{j^-}x}\phi_i(x)\prod_{i=1}^n {\boldsymbol{\Psi}}_\text{vac}[\phi_{i,A}, \phi_{i,{\bar{A}}}]\,,
\end{align}
\begin{align}
&\prod_{i=1}^n {\boldsymbol{\Psi}}_{\{p_{j^+}p_{j^-}\}}[\phi_{i+1,A}, \phi_{i,{\bar{A}}}]^* =\left(\prod_{j^+}\frac{2E_{p_{j^+}}}{L}\prod_{j^-}\frac{2E_{p_{j^-}}}{L}\right)^\frac{n}{2}
\\ &\times \prod_{i=1}^n\prod_{j^+}\left(\int_A\mathrm{d}x e^{-ip_{j^+}x}\phi_{i+1}(x)+ \int_{\bar{A}}\mathrm{d}x e^{-ip_{j^+}x}\phi_i(x)\right)\\ &\times \prod_{j^-}\left(\int_A\mathrm{d}x e^{-ip_{j^-}x}\phi^\dagger_{i+1}(x)+ \int_{\bar{A}}\mathrm{d}x e^{-ip_{j^-}x}\phi^\dagger_i(x)\right)\prod_{i=1}^n {\boldsymbol{\Psi}}_\text{vac}[\phi_{i+1,A}, \phi_{i,{\bar{A}}}]\,.
\end{align}
Putting everything together we have that
\be
\label{ratio_of_traces_zero_flux}
\frac{\Tr_A\left(\rho^n_{A}\right)}{\Tr_A(\rho^n_{0,A})} = \left(\prod_{j^+}\frac{2E_{p_{j^+}}}{L}\prod_{j^-}\frac{2E_{p_{j^-}}}{L}\right)^n \bra \prod_{i=1}^n \prod_{j^+} U^+_i(p_{j^+})V^+_i(p_{j^+}) \prod_{j^-} U^-_i(p_{j^-})V^-_i(p_{j^-})\ket 
\ee
where the correlation function is defined with respect to the Gaussian measure:
\be
\label{Gaussian_correlator}
\bra \mathcal{O}[\phi_1,\phi^\dagger_1\dots,\phi_n, \phi^\dagger_n]\ket = \frac{\int \mathcal{D}\phi_1\dots \mathcal{D}\phi_n \mathcal{O}[\phi_1,\phi^\dagger_1\dots,\phi_n, \phi^\dagger_n] \exp\left[-\mathcal{G}\right]}{\int \mathcal{D}\phi_1\dots \mathcal{D}\phi_n \exp\left[-\mathcal{G}\right]}
\ee
and the operators are
\be
\label{U_uncharged}
U_i^+(p) = \int_{A \bigcup {\bar{A}}}\mathrm{d}x\, e^{ipx} \phi^\dagger_i(x) \quad , \quad U_i^-(p) = \int_{A \bigcup {\bar{A}}}\mathrm{d}x \,e^{ipx} \phi_i(x)
\ee
\begin{align}
\label{V_uncharged}
V_i^+(p) &= \int_A\mathrm{d}x\, e^{-ipx} \phi_{i+1}(x) + \int_{\bar{A}}\mathrm{d}x \,e^{-ipx} \phi_i(x)\,, \\
V_i^-(p) &= \int_A\mathrm{d}x\, e^{-ipx} \phi^\dagger_{i+1}(x) + \int_{\bar{A}}\mathrm{d}x \,e^{-ipx} \phi^\dagger_i(x)\,.
\end{align}
If the excited state is of the form \eqref{multi-particle state_equal_momenta}, the result \eqref{ratio_of_traces_zero_flux} is minimally modified. The terms inside the correlator are exactly the same, except that now the range of the indices $j^\pm$ is that specified after \eqref{multi-particle state_equal_momenta} and the prefactor is modified to:
\be
\left(\prod_{j^+=1}^{m^+}\frac{1}{k_{j^+}^+!}\left(\frac{2E_{p_{j^+}}}{L}\right)^{k_{j^+}^+}\prod_{j^-=1}^{m^-}\frac{1}{k_{j^-}^-!}\left(\frac{2E_{p_{j^-}}}{L}\right)^{k_{j^-}^-}\right)^n\,.
\ee

\subsection{non-trivial flux insertion}
We now come to the quantity \eqref{ratiof} with $\alpha \ne 0$, assuming the excited state to be of the form \eqref{multi-particle state}. Let us consider the denominator first. Because of the flux insertion to the right of the $n$th operator $\rho_{0,A}$, everything is the same as in \eqref{denominator_no_flux} except for the last resolution of the identity, which produces a term:
\be 
\bra \Psi | \phi_{n {\bar{A}}} \ket e^{2\pi i \alpha Q_A} |\phi_{1 A} \ket = \bra \Psi | e^{2\pi i \alpha}\phi_{1 A}, \phi_{n {\bar{A}}} \ket = {\boldsymbol{\Psi}}[\phi_{1 A}e^{2\pi i \alpha}, \phi_{n {\bar{A}}}]^*\,.
\ee
Thus, the Gaussian measure in the presence of the charge is modified as follows:
\be
\Tr_A(\rho^n_{0,A}e^{2\pi i \alpha Q_A}) = \int \mathcal{D}\phi_1\dots \mathcal{D}\phi_n \exp \left[-\mathcal{G}_\alpha\right]\,,
\ee
with
\begin{align}
\label{continuous_matrix_alpha}
\mathcal{G}_\alpha = \sum_{i=1}^n &\left[2\left(\int_{\substack{x \, \in \, A \\ y \, \in \, A}} + \int_{\substack{x \, \in \, {\bar{A}} \\ y \, \in \, {\bar{A}}}}\right)\phi_i^\dagger(x)K(x-y)\phi_i(y)  \right. \nonumber \\&+ \left. \left(\int_{\substack{x \, \in \, A \\ y \, \in \, {\bar{A}}}} \left(\phi^\dagger_i(x) + \phi^\dagger_{i+1}(x) e^{-2\pi i \alpha \delta_{i,n}}\right) K(x-y)\phi_i(y) + \text{c.c.}\right) \right]\,.
\end{align}
As for the numerator of (\ref{ratiof}), we have similarly
\begin{align}
\Tr_A(\rho^n_{A}) = \int \mathcal{D}\phi_1\dots \mathcal{D}\phi_n &\prod_{i=1}^{n-1}{\boldsymbol{\Psi}}_{\{p_{j^+},p_{j^-}\}}[\phi_{i A}, \phi_{i {\bar{A}}}]{\boldsymbol{\Psi}}_{\{p_{j^+},p_{j^-}\}}[\phi_{i+1,A},\phi_{i {\bar{A}}}]^* \nonumber \\ &\times{\boldsymbol{\Psi}}_{\{p_{j^+},p_{j^-}\}}[\phi_{n A}, \phi_{n {\bar{A}}}]{\boldsymbol{\Psi}}_{\{p_{j^+},p_{j^-}\}}[\phi_{1,A}e^{2\pi i \alpha},\phi_{n {\bar{A}}}]^*\,.
\end{align}
The product in the first line gives what we already had found in the $\alpha=0$ case, with the replica index running up to $n-1$ only. 
The product of the last two functionals gives:
\begin{align}
&\left(\prod_{j^+}\alpha_{p_{j^+}}[\phi^\dagger_n]\alpha_{p_{j^+}}[\phi^\dagger_{1A}e^{-2\pi i \alpha}, \phi^\dagger_{n{\bar{A}}}]^*\prod_{j^-}\beta_{p_{j^-}}[\phi_n]\beta_{p_{j^-}}[\phi_{1A}e^{2\pi i \alpha}, \phi_{n{\bar{A}}}]^*\right)\nonumber \\&\times{\boldsymbol{\Psi}}_\text{vac}[\phi_{n A}, \phi_{n {\bar{A}}}]{\boldsymbol{\Psi}}_\text{vac}[\phi_{1,A}e^{2\pi i \alpha},\phi_{n {\bar{A}}}]^*\,.
\end{align}
and from the definitions of the $\alpha$ and $\beta$ operators we obtain our final result:
\begin{align}
\label{ratio_of_charged_moment_functionals}
\frac{\Tr_A\left(\rho^n_{A}\,e^{2\pi i \alpha Q_A}\right)}{\Tr_A\left(\rho^n_{0,A}\,e^{2\pi i \alpha Q_A}\right)} &= \left(\prod_{j^+}\frac{2E_{p_{j^+}}}{L}\prod_{j^-}\frac{2E_{p_{j^-}}}{L}\right)^n \nonumber\\&\times\langle \prod_{i=1}^n \prod_{j^+} U^+_i(p_{j^+})V^+_i(p_{j^+}) \prod_{j^-} U^-_i(p_{j^-})V^-_i(p_{j^-})\rangle_\alpha\,,
\end{align}
where $\bra \dots \ket_\alpha$ is defined as in \eqref{Gaussian_correlator} with respect to the Gaussian measure (\ref{continuous_matrix_alpha}). 
 $U_i^\pm(p)$ are defined exactly as in \eqref{U_uncharged}, while $V_i^\pm(p)$ are now modified by the presence of the charge as follows:
    \begin{align}
    \label{V_charged}
    V_i^+(p) &= e^{2\pi i \alpha \delta_{i,n}}\int_A\mathrm{d}x\, e^{-ipx} \phi_{i+1}(x) + \int_{\bar{A}}\mathrm{d}x \,e^{-ipx} \phi_i(x)\,, \\
    V_i^-(p) &= e^{-2\pi i \alpha \delta_{i,n}}\int_A\mathrm{d}x\, e^{-ipx} \phi^\dagger_{i+1}(x) + \int_{\bar{A}}\mathrm{d}x \,e^{-ipx} \phi^\dagger_i(x)\,.
    \end{align}
If the excited state is given by \eqref{multi-particle state_equal_momenta}, the right-hand side of \eqref{ratio_of_charged_moment_functionals} is changed exactly as in the case $\alpha = 0$.


\end{document}